\documentclass[pra,twocolumn,superscriptaddress,nofootinbib, amsmath,amssymb, aps]{revtex4-2}
\usepackage{graphicx}
\usepackage{bm, amsmath, amssymb}
\usepackage{color}
\usepackage{soul}
\usepackage[T1]{fontenc}
\usepackage{amstext}
\usepackage{amsfonts}

\usepackage{ulem}
\usepackage{xcolor}
\usepackage{float}
\usepackage{algorithm}
\usepackage{algpseudocode}
\usepackage{multirow}
\usepackage{physics}
\usepackage{hyperref}
\usepackage{xr}
\usepackage{cleveref}
\newtheorem{definition}{Definition}[section]

\def\BibTeX{{\rm B\kern-.05em{\sc i\kern-.025em b}\kern-.08em
    T\kern-.1667em\lower.7ex\hbox{E}\kern-.125emX}}
\begin{document}

    \title{Geometric correspondence of noisy quantum dynamics and universal robust quantum gates}
\author{Yong-Ju Hai}
\affiliation{Shenzhen Institute for Quantum Science and Engineering (SIQSE), Southern University of Science and Technology, Shenzhen, P. R. China}
\affiliation{Department of Physics, Southern University of Science and Technology, Shenzhen 518055, China}
\author{Yao Song}
\affiliation{Shenzhen Institute for Quantum Science and Engineering (SIQSE), Southern University of Science and Technology, Shenzhen, P. R. China}
\author{Junning Li}
\affiliation{Department of Physics, City University of Hong Kong, Tat Chee Avenue,
Kowloon, Hong Kong SAR, China}
\author{Junkai Zeng}
\affiliation{Shenzhen Institute for Quantum Science and Engineering (SIQSE), Southern University of Science and Technology, Shenzhen, P. R. China}
\author{Xiu-Hao Deng}
\email{dengxh@sustech.edu.cn}
\affiliation{Shenzhen Institute for Quantum Science and Engineering (SIQSE), Southern University of Science and Technology, Shenzhen, P. R. China}
\affiliation{International Quantum Academy (SIQA), and Shenzhen Branch, Hefei National Laboratory, Futian District, Shenzhen, P. R. China}

\begin{abstract}

Quantum information processing faces a significant hurdle: noise. Different noise sources induce varying errors in quantum operations depending on the underlying dynamics. To gain a deeper understanding of these error mechanisms, we introduce the concept of quantum error evolution diagrams (QEEDs). QEEDs establish a dual correspondence between driven noisy quantum dynamics and geometric space curves, providing quantitative geometric metrics to assess the severity of the noises. This framework enables the design of universal robust quantum gates that correct errors induced by generic noise. Additionally, we present a protocol for constructing a universal set of single- and two-qubit robust quantum gates with simple and smooth control pulses of arbitrary length and fidelities exceeding 99.99\% across a wide range of noise strengths. Our work provides new insights into the geometric nature of noisy quantum dynamics and paves the way for developing strategies to dynamically correct quantum errors.


\end{abstract}

\maketitle
\section{Introduction}

Paving the way to the future of quantum technologies and quantum computing hinges upon our ability to exercise precise and robust control over inherently noisy quantum systems. \cite{arute2019quantum, wu2021strong, krinner2022realizing, preskill2022physics, preskill2018quantum, glaser2015training, koch2016controlling, koch2022quantum}. Recent progress has pushed the quality of quantum gates to approach the fault-tolerant threshold in isolated characterizations \cite{barends2014superconducting, xue2022quantum, ballance2016high,bluvstein2022quantum}; yet, realistic multi-qubit systems are subject to a myriad of noises coupling to all system operators, precipitating quantum decoherence and gate errors \cite{reinhold2020error,deng2021correcting}. Remarkably, these noises do not invariably breed errors in quantum operations driven by evolution under specific control fields. Robust quantum gates, critical in integrating a large array of qubits \cite{xiang2020simultaneous,terhal2020towards, cai2021bosonic,manovitz2022trapped,zhou2023quantum}, have been fashioned to dynamically correct errors \cite{khodjasteh2009dynamically,zeng2018general,zeng2019geometric,barnes2022dynamically}. Conventionally, these gates are synthesized through pulse-shaping techniques that exploit numerical optimization, a method increasingly constrained by computational cost \cite {khaneja2005optimal,caneva2011chopped,yang2019silicon,figueiredo2021engineering,ribeiro2017systematic,song2022optimizing}, and potentially inapplicable to arbitrary noise models or scalable multi-qubit systems. Alternatives employ the geometric phase accrued via cyclic quantum evolution, achieving immunity to specific noise types but at the cost of slower gates \cite {berry1990geometric,zu2014experimental,balakrishnan2004classical,ekert2000geometric,xu2020experimental}. To better engineer the control field at the pulse level, perturbative treatment of various noises has been utilized to study noisy quantum dynamics \cite{green2013arbitrary}, leading to the proposal of dynamically-corrected gates designed to manage dephasing noises by mapping evolution onto geometric curves \cite{zeng2018general,zeng2019geometric,barnes2022dynamically}. Nonetheless, both geometric methods grapple with significant challenges and practical constraints, including the limitation to a single error in a single qubit  \cite{zeng2018general}. Realistic systems are subject to various noises that couple to all system operators. They may arise from inaccurate chip characterization, disturbances during quantum control, scalability issues due to excessive control lines, unwanted interactions, crosstalk, etc. \cite{google2023suppressing}. Thus, a robust quantum gate's resilience to such noise combinations becomes crucial. Therefore, our focus must be on comprehending the genesis of errors in noisy quantum dynamics and the control field's error correction mechanisms within this noise-laden backdrop.

\begin{figure*}
\centering
\includegraphics[width=0.9\textwidth]{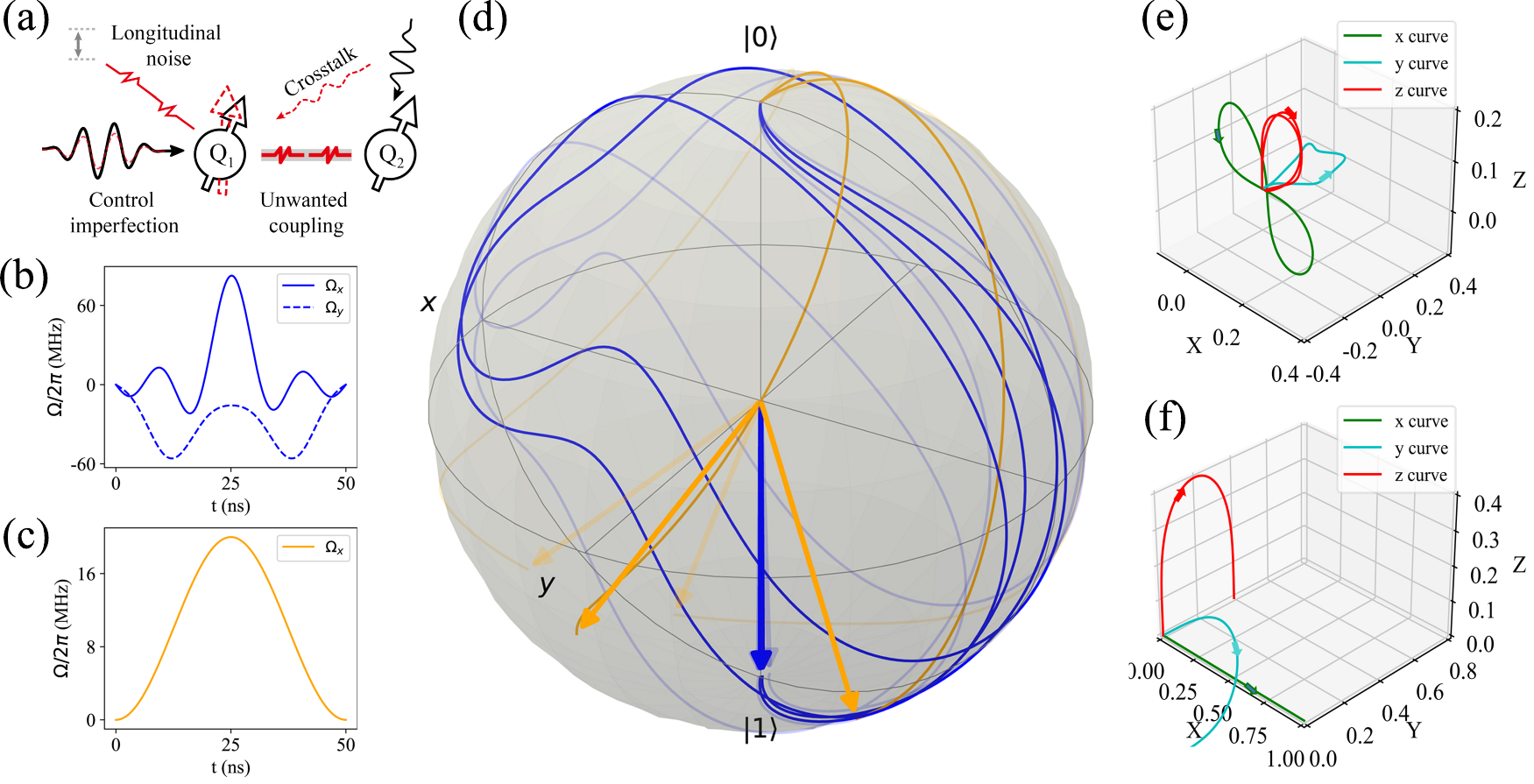}
\caption{The illustration of geometric correspondence between noisy driven quantum dynamics and the corresponding QEEDs. (a) shows some types of noises in a multi-qubit system. (b) and (c) show control pulses for an $X_\pi$ gate. The blue shows the robust control pulse (RCP) introduced in this manuscript, while the orange is Gaussian. (d) illustrates the quantum evolution driven by the two pulses (with corresponding colors) on the Bloch sphere subject to randomly generated noises. The pulses are intended to implement an $X_\pi$ gate, but orange trajectories end up at different states on the Bloch sphere, while the RCP drives the blue trajectories to the same spot despite the different paths because of the noises. (e) and (f) are the QEEDs which illustrate the error curves corresponding to the RCP and Gaussian pulses, respectively. }
\label{Fig_schematics}
\end{figure*}

In this manuscript, we establish the essential correspondence between the noisy quantum dynamics and multiplex space curves that form a diagram to analyze control errors, which we call the quantum error evolution diagrams (QEED). These curves are parametrized by generalized Frenet-Serret frames and thus enable a bijective map with the Hamiltonian. We use the multiplex curve diagrams to characterize the noisy quantum evolution quantitatively. We show that the diagrams provide quantitative metrics of the control robustness to different noises. We derive the necessary and sufficient conditions for general robust control. To make this theoretical framework applicable for robust control in experiments, we develop a simple and systematic protocol to find arbitrary robust control pulses with the simplest waveforms, consisting of only a few Fourier components. This protocol could be automated given only the system parameters, making it convenient for experiments. We demonstrate that these pulses could have arbitrary gate time while preserving the same robustness. We use these pulses to implement universal robust quantum gates. We use realistic multi-qubit models of semiconductor quantum dots and superconducting transmons for numerical simulations. We obtain high gate-fidelity plateaus above the fault-tolerant threshold over a wide range of noise strength. These results demonstrate the effectiveness of our framework and suggest its potential applications. Our work provides a new perspective on noisy quantum dynamics and their relation to robust control by revealing an essential geometric correspondence between them. Our work also offers a practical solution for engineering universal robust quantum gates with simple pulses that are ready for experiments.

The paper is organized as follows: In Sec.~\ref{Sec_theory}, we present our analytical theory based on QEED; in Sec.~\ref{Sec_protocol} we develop our protocol for finding robust control pulses; in Sec.~\ref{Sec_application}, we apply our protocol to realistic models of spin qubits and transmons and present the outstanding results of robust gates; in Sec.~\ref{Sec_Conclude}, we discuss some extensions and limitations of our approach and then conclude with some remarks. Some details about the derivation and numerical results are presented in the Appendix.

\section{Analytical theory}
\label{Sec_theory}

\subsection{Model settings}

A noisy quantum system's dynamics could be modeled with a stochastic Schrödinger equation  $i\hbar \dot \psi=H(\{\delta_i(t)\};t)\psi$, which is a variant of the traditional Schrödinger equation that incorporates randomness or noise into the quantum dynamics of a system. To clarify the meaning of the discussed terms, we begin with the following definitions:

\begin{definition}{Noise:}
The stochastic variable $\delta_i(t)$ is designated as "Noise" within this manuscript.
\end{definition}
The time-dependence of $\delta_i(t)$ is contingent upon the origins of noise, including spectrum broadening in ensembles, the presence of crosstalk in multi-qubit systems, parameter drifts, unwanted couplings \cite{deng2021correcting,krinner2020benchmarking}, among other factors. In general, $\delta_i(t)$ exhibits time-dependent characteristics and fluctuates randomly within a specific noise spectrum in a given system.

\begin{definition}{Error:}
The error is the result of the noisy evolution. The full evolution operator $U$ contains the ideal part $U_0$ and the error $U_\epsilon$, i.e. $U=U_0 U_\epsilon$.
\end{definition}

Our manuscript focuses on a geometric perspective of how the error \cite{gottesman1997stabilizer} evolves and how it is corrected dynamically. 

The generic Hamiltonian can be decomposed as:
\begin{equation}
H(\{\delta_i(t)\};t)=H_{0}+H_{c}(t)+V(\{\delta_i(t)\}).  \label{Eq_Hfull}
\end{equation}%
Here, the system Hamiltonian $H_{0}$ added with the control term $H_{c}(t)$ drives the desired time evolution. And $V=\boldsymbol{\delta }\cdot\hat{\mathbf{O}}$ is the noise term, where vector $\boldsymbol{\delta }$'s components are noises $\delta_i(t)$ \cite{carballeira2021stochastic,clerk2022introduction} that couple to the target quantum system via the operators $\hat{\mathbf{O}}=\{O_{1},...,O_{n}\}$. In a $SU(2)$ system, $\hat{\mathbf{O}}=\hat{\boldsymbol{\sigma}}$ which are Pauli operators.

The idea of robust control using geometric correspondence is illustrated in Fig.~\ref{Fig_schematics}. A trivial control field $H_{c}(t)$, for example, a cosine pulse, supposedly drives the raw system $H_{0}$ to a target final state $|\psi _{f}\rangle $. But subject to the noise $V$, the system's evolution deviates from the ideal trajectory $U_{0}$ and undergoes a noisy evolution $U = U_0 U_\epsilon$. The error final state $|\psi _{e}\rangle $ has an unknown distance from the desired final state depending on the noise strength. To correct the errors, control Hamiltonian $H_{c}(t)$ formed with robust control pulse (RCP) obtained by geometric correspondence should drive the system to the expected state $|\psi _{f}\rangle $, disregarding the presence of unknown noises. This means that at some time $T_g $, $U(T_g)=U_0$ and $U_\epsilon(T_g)=I$. Nonetheless, solving the RCP is very challenging. We will show how to formulate error measures of control pulses and construct arbitrary RCPs, by establishing the geometric correspondence between space curves and the noisy quantum dynamics.

In order to obtain analytical solutions, our theory is formulated on the two-state systems, giving the SU(2) dynamics. The dynamics of higher dimensional systems correspond to more complex geometric structures and are beyond the scope of this manuscript. However, some complex system's dynamics could be decomposed into direct sums or products of $SU(2)$ operators; then the work presented in this manuscript still applies. The raw Hamiltonian for a single two-state Hilbert space $\text{span}\{\left\vert a\right\rangle ,\left\vert b\right\rangle \}$ is $H_{raw}=-\frac{\omega }{2}\sigma _{z}$. If driven transversely with a control field, it is usually written in the rotating frame with the control field
\begin{equation}
H_{0}^{rot}=-\frac{\Delta }{2}\sigma _{z},
\end{equation}%
where $\Delta$ is the detuning between qubit frequency and control field. The formulas in this manuscript are all present in the rotating frame, so in the following, we will drop the superscript "rot". The control and the noise Hamiltonian could be written in the general form 
\begin{eqnarray}
H_{c}(t) &=&\Omega _{z}(t)\sigma _{z}+\Omega _{x}(t)\sigma _{x}+\Omega_{y}(t)\sigma _{y}  \label{Eq_Hcontrol} \\
V &=&\delta _{z}\sigma _{z}+\delta _{x}\sigma _{x}+\delta _{y}\sigma _{y},
\label{Eq_GendeltaH}
\end{eqnarray}%
where the Pauli matrices $\sigma _{x}=|a\rangle \langle b|+|b\rangle \langle a|$, $\sigma_{y}=-i|a\rangle \langle b|+i|b\rangle \langle a|$, $\sigma _{z}=|a\rangle\langle a|-|b\rangle \langle b|$. 
The longitudinal noise $\delta_z$ has many origins including variations in frequency or fluctuating spectral splitting, which could originate from the coupling to other quantum systems, e.g., the neighboring qubits, two-level defects, or quantum bath \cite{burnett2019decoherence,lisenfeld2019electric}. On the other hand, the transverse noises $\delta _{x}$ and $\delta _{y}$ result from energy exchange with the environment, such as relaxation, crosstalk, and imperfect control. The combination of the noises coupled to the three Pauli bases gives the generic form of noises and hence a general error model, regardless of specific forms of error. In this manuscript, the formulation of the theory assumes that the noise fluctuation time scale is longer than a quantum gate. So naturally the noise is in this quasi-static limit and the vector $\boldsymbol{\delta}$ becomes a constant vector.

Based on this assumption, we will show how to engineer the control Hamiltonian to correct the errors due to noises in its perpendicular directions. And hence for generic noises in the $x,y,z$ directions we need controls in at least two directions, i.e., two terms out of the three in the control Hamiltonian in Eq.~(\ref{Eq_Hcontrol}) are sufficient to achieve dynamical error suppression. Specifically, in the following discussion, we let 
\begin{equation}
H_{c}(t)=\frac{\Omega (t)\cos \Phi (t)}{2}\sigma _{x}+\frac{\Omega (t)\sin
\Phi (t)}{2}\sigma _{y},  \label{Eq_XYdrive}
\end{equation}%
where $\Omega (t)$ and $\Phi (t)$ are the control amplitude and modulated phase of the transverse control field.

\subsection{Geometric correspondence}
\label{subsecGeocorresp}

This study elucidates the fundamental geometric relationship between driven noisy quantum dynamics and space curves, upon which we further establish constraints for robust control. The comprehensive time evolution is described by $U(t)=\mathcal{T}\exp\{-i\int_{0}^{t}H(u )du \}$, which is generated by the complete noisy Hamiltonian $H(t)$ as depicted in Eq.(\ref{Eq_Hfull}). In contrast, the evolution of noiseless driven dynamics is characterized by $U_{0}(t)=\mathcal{T}\exp \{-i\int_{0}^{t}(H_{0}+H_{c}(u))du \}$. In the context of the interaction picture, represented by $U_{0}(t)$, the noise Hamiltonian undergoes a transformation to $V_{I}=U_{0}^{\dagger }VU_{0}=U_{0}^{\dagger }(\boldsymbol{\delta }\cdot \hat{\boldsymbol{\sigma}})U_{0}$. As explained in the previous section, the model assumes time-independent noises, thereby rendering $\boldsymbol{\delta }$ a constant within the time integral. For the $j$-component of the noise source, the transformation of its Pauli operator $\sigma_j$ is parameterized by a displacement vector $U_0^\dagger \sigma_j U_0 = d\boldsymbol{r}^{j}(t)\cdot \hat{\boldsymbol{\sigma }}$, with its time derivative defining a velocity $\boldsymbol{T}_{j}(t)=\dot{\boldsymbol{r}}^{j}(t)$. As denoted by $||\boldsymbol{T}_{j}\cdot \hat{\boldsymbol{\sigma }}||_F=\sqrt{2}||\boldsymbol{T}_j||=||U_{0}^{\dagger }(t)\sigma _{j}U_{0}(t)||_F=\sqrt{2}$, $\boldsymbol{T}_{j}$ serves as a unit vector, indicating that the $j$-error motion proceeds at a constant velocity. Therefore, the Hamiltonian exhibits a geometric correspondence 
\begin{align}
V_{I}(t)& =\delta _{x}\mathbf{T}_{x}(t)\cdot \hat{\boldsymbol{\sigma }}%
+\delta_{y}\mathbf{T}_{y}(t)\mathbf{\cdot }\hat{\boldsymbol{\sigma }}+\delta _{z}\mathbf{T}_{z}(t)\cdot \hat{\boldsymbol{\sigma }}
\notag \\
& =\sum_{j,k=x,y,z}\delta _{j}T_{jk}(t)\sigma _{k}.
\end{align}%
Here, $T_{jk}(t)$ serves as a tensor that links the $j$-component of the noise source $\delta _{j}$ with the Pauli term $\sigma _{k}$.

In the decomposition $U(t)=U_{0}(t)U_{\epsilon}(t)$, $U_{\epsilon}(t)$ can be identified as the \textit{error evolution} that pertains to the deviation observed between the noisy $U(t)$ and the noiseless $U_{0}(t)$ and is generated by interaction picture noise Hamiltonian following the integral in time order as 
\begin{align}
U_{\epsilon}(t)& =\mathcal{T}\exp \{-i\int_{0}^{t}du V_{I}(u)\}  \notag \\
& =\mathcal{T}\exp \{-i\sum_{j=x,y,z}\delta _{j}\int_{0}^{t}du \lbrack 
\mathbf{T}_{j}(u)\cdot \hat{\boldsymbol{\sigma }}\mathbf{]}\}  \notag
\\
& =\mathcal{T}\exp \{-i\sum_{j=x,y,z}\delta _{j}\mathbf{r}^{j}%
(t)\cdot \hat{\boldsymbol{\sigma }}\}  \label{Eq_UItimeordering}.
\end{align}%
Therefore, the error dynamics could be described by the kinematics of a moving point. The displacement $\mathbf{r}^{j}(t)$ sketches a space curve in $\mathbb{R}^{3}$, corresponding to the noise coupled to $\sigma _{j}$. So we call $\mathbf{r}^{j}(t)$ the \textit{$j$-error curve}. Therefore, the error evolution $U_{\epsilon}(t)$ could be described by three error curves, forming a diagram to describe the error dynamics of the two-state system. In this manuscript, we call it a Quantum Error Evolution Diagram (QEED). It is known that any continuous, differentiable space curve could be defined necessarily and efficiently with the Frenet-Serret frame \cite{o2006elementary, pressley2010elementary} with a tangent, a normal, and a binormal unit vector $\{\mathbf{T},\mathbf{N},\mathbf{B}\}$. Moreover, there is a correspondence between the curve length and the evolution time $%
L(t)=\int_{0}^{t}du ||\mathbf{T}_{j}(u)\mathbf{||}=\int_{0}^{t}du =t$. Using the $j$-error curve and its tangent vector defined above, the explicit geometric correspondence could be established.

As an example, for the control Hamiltonian Eq.~(\ref{Eq_XYdrive}) and noise in $z$-direction, $H=\delta _{z}\sigma _{z}+H_{c}(t)$. Following the derivation in the Appendix.~\ref{AppendixA}, the explicit geometric correspondence between the control Hamiltonian and the error curve is given by 
\begin{equation}
\begin{array}{c}
\mathbf{T}\cdot \hat{\boldsymbol{\sigma }}\mathbf{=}U_{0}^{\dagger
}(t)\sigma _{z}U_{0}(t) \\ 
\mathbf{N}\cdot \hat{\boldsymbol{\sigma}}=U_{0}^{\dagger }(t)(-\sin \Phi
(t)\sigma _{x}+\cos \Phi (t)\sigma _{y})U_{0}(t) \\ 
\mathbf{B}\cdot \hat{\boldsymbol{\sigma}}=U_{0}^{\dagger }(t)(-\cos \Phi
(t)\sigma _{x}-\sin \Phi (t)\sigma _{y})U_{0}(t).%
\end{array}
\label{Eq_z_correspondence1}
\end{equation}%
Combining Eq.~(\ref{Eq_XYdrive}) and Eq.~(\ref{Eq_z_correspondence1}) we get the relation between the Frenet vectors and the control Hamiltonian  
\begin{equation}
\left\{ 
\begin{array}{c}
\kappa (t)=\dot{\mathbf{T}}\cdot \mathbf{N}%
=\Omega (t) \\ 
\tau (t)=-\dot{\mathbf{B}}\cdot \mathbf{N}=\dot{%
\Phi}(t),
\end{array}
\right.   \label{Eq_z_correspondence2}
\end{equation}%
where $\kappa (t)$ and $\tau (t)$ are, respectively,  the signed curvature and the singularity-free torsion of $z$-error curve. 

So far, we have established the geometric correspondence between the noisy quantum dynamics and the kinematic properties of space curves. This correspondence is a bijective map which means that given either a Hamiltonian or a curve, its counterpart could be solved straightforwardly via Eq.~(\ref{Eq_z_correspondence1}) or Eq.~(\ref{Eq_z_correspondence2}). We refer to Appendix.~\ref{AppendixA} and \ref{AppendixC} for more details.

\subsection{Robust control}
\label{subsecRobCon}

Intuitively, a control is called robust if the control precision in the presence of noises remains almost the same as in the noiseless case $\delta=0$. 
\begin{definition}{Robust quantum gate:}
If for $\forall \varepsilon$,  there exists a $\delta_0>0$, for $\forall \delta \in [ 0,\delta_0)$, $\frac{\partial \parallel U_\epsilon-I\parallel}{\partial\delta} < \varepsilon$, we call the (time evolution) unitary operator robust against $\delta$. The quantum gate generated by this unitary operator is called a robust quantum gate. 
\end{definition}

The robust control of the two-state system requires the error evolution to vanish while driving the system to achieve a target gate at a specific gate time $T_g $, i.e., satisfying the robust condition $U(T_g)=U_{0}(T_g )$ and $U_{\epsilon}(T_g )=I$. To obtain the explicit form of $U_{\epsilon}$ in terms of the error curve, we use the equivalency between the time ordering representation and the Magnus expansion \cite{blanes2009magnus} to
obtain 
\begin{align}
U_{\epsilon}(t)& =\exp \{-i\sum_{n}[(\mathbf{\delta }\cdot )^{n}%
\hat{\mathbf{A}}_{n}(t)]\}  \notag \\
& =\exp \{-i[\delta _{j}A_{1}^{j}+\delta _{j}\delta _{k}A_{2}^{jk}+O(\delta
^{3})]\},
\end{align}%
where $\hat{\mathbf{A}}_{n}$ is $n$th-order tensor corresponding
to $n$th-order of the Magnus series and the Einstein summation is used. We have also assumed that all the $\delta _{x,y,z}$ terms are at the same perturbative order. The exponential form can also be further expanded to polynomial series 
\begin{equation}
U_{\epsilon}(t)=I-i\delta _{j}A_{1}^{j}-[\frac{1}{2}(\delta
_{j}A_{1}^{j})^{2}+i\delta _{j}\delta _{k}A_{2}^{jk}]+O(\delta ^{3}),
\label{Eq_perturbative}
\end{equation}%
where %
\begin{equation}
\left\{ 
\begin{array}{c}
A_{1}^{j}(t)=\int_{0}^{t }du(\mathbf{T}_{j}\cdot \hat{\boldsymbol{\sigma }}\mathbf{)} \\ 
A_{2}^{jk}(t)=\frac{1}{2}\int_{0}^{t}du \lbrack \dot{A}%
_{1}^{j}(t),A_{1}^{k}(t)] \\ 
A_{n+1}^{jkl...}(t)=\frac{1}{2}\int_{0}^{t}du \lbrack \dot{A}%
_{n}^{j}(t),A_{n}^{kl...}(t)].
\end{array}%
\right.  \label{Eq_perturbOrders}
\end{equation}

Utilizing the geometric correspondence present above, the robust constraints could be established up to arbitrary perturbative orders. Correcting the leading-order error at time $T_g $ requires the first-order term $A_{1}^{j}(t)$ in Eq.~(\ref{Eq_perturbative}) to vanish. Using $\mathbf{T}_{j}(t)=\dot{\mathbf{r}}^{j}(t)$, we get an explicit
geometric representation of the error evolution 
\begin{equation}
A_{1}^{j}(t)=\mathbf{r}^{j}(t)\cdot \hat{\boldsymbol{\sigma }}.
\label{Eq_1order}
\end{equation}%
The geometric correspondence of $A_{1}^{j}(t)$ is given by the displacement $\mathbf{r}^{j}(t)$ of the $j$-error curve. Therefore, the condition of control robustness up to the leading order is%
\begin{equation}
\mathbf{r}^{j}(T_g )=0.
\end{equation}
We thus have $\mathbf{r}^{j}(0)=\mathbf{r}^{j}(T_g ) = 0$ and the curve is closed during the gate time.

Eq.~(\ref{Eq_perturbOrders}) infers that the higher-order terms contain more complex geometric properties. For simplicity, we consider the case when the error lies in only one axis $j$. The second-order term in Eq.~(\ref{Eq_perturbative}) becomes
\begin{equation}
A_{2}^{jj}(t)=i\int_{0}^{t}\dot{\mathbf{r}}^{j}(u
)\times \mathbf{r}^{j}(u )du \cdot \hat{\boldsymbol{\sigma}%
}=i\mathbf{R}^{j}(t)\cdot \hat{\boldsymbol{\sigma}},
\label{Eq_2order}
\end{equation}%
now has a geometric meaning that $\mathbf{R}^{j}(t)=\int_{0}^{t}%
\dot{\mathbf{r}}^{j}(u )\times \mathbf{r}^{j}%
(u )du $ forms directional integral areas on $y$-$z$, $z$-$x$, and $x$-$y$ planes enclosed by the projections of the space curve. Similarly, the second-order robustness conditions require $\frac{1}{2}(\delta_{j}A_{1}^{j})^{2}+i\delta _{j}^{2}A_{2}^{jj}=0$, i.e.,
\begin{equation}
\left\{ 
\begin{array}{c}
\mathbf{r}^{j}(T_g )=0 \\ 
\mathbf{R}^{j}(T_g )=0.
\end{array}%
\right.   \label{Eq_2robust}
\end{equation}%
Higher-robustness conditions also refer to the vanishing net areas of the corresponding space curves only with more closed loops \cite{barnes2022dynamically}. Nonetheless, higher-order robustness means more constraints so that the search for control pulses becomes more challenging, and the resulting control pulses are typically longer and more complicated, making experimental realization infeasible. Therefore, aiming at robustness for orders higher than two is usually unnecessary. We consider only the constraints up to leading terms in the pulses construction protocol presented in the following Section~\ref{Sec_protocol}.

To explain the assumption addressed at Eq.~(\ref{Eq_XYdrive}), we now present theorems to answer the questions: 1. What are the necessary conditions to correct the errors? 2. What are the sufficient conditions to correct the errors? 

\textbf{Theorem:} (\textit{Non-correctable condition}) If $[V,H_{c}(t)]=0$ for $\forall t\in \lbrack 0,T _{g}]$, the error evolution cannot be dynamically corrected. Specifically, if $[\sigma _{j},H_{c}(t)]=0$, the error generated by $j$ noise cannot be dynamically corrected by $H_{c}(t)$.

Proof. Using the geometric correspondence, the proof for this theorem becomes explicit and intuitive. A necessary condition for the dynamical correctability is that the velocity $\mathbf{T}$ of the error evolution could be modified by the control Hamiltonian, namely $\mathbf{N}\neq 0$ at some time $t\in\lbrack 0,T_{g}]$. Whether a trajectory $\mathbf{r}$ is curved or not depends on its normal vector $\mathbf{N}$, corresponding to the dependence of error evolution on $H_{c}(t)$. From the geometric correspondence introduced above, we know%
\begin{align}
\mathbf{N}(t)\cdot \hat{\boldsymbol{\sigma}}& =\frac{d(U_{0}^{\dagger }VU_{0})}{dt}  \notag
\\
& =\dot{U}_{0}^{\dagger }VU_{0}+U_{0}^{\dagger }\dot{%
V}U_{0}+U_{0}^{\dagger }V\dot{U}_{0}  \notag \\
& =U_{0}^{\dagger }[H_{c},V]U_{0}.
\end{align}%
Here for quasi-static noise, $\dot{V}=0$ within the gate time. Therefore, if $[V,H_{c}(t)]=0$ for $\forall t\in \lbrack 0,T _{g}]$, $\mathbf{N}(t)\equiv 0$. In the geometric frame, it means that the error curve remains in the same direction, and hence the error evolution cannot be dynamically corrected.

Specifically, for “$j$ noise, $V_{j}=\delta _{j}\sigma _{j}$, the non-correctable condition $[V,H_{c}(t)]=0$ becomes $[\sigma _{j},H_{c}(t)]=0$. Q.E.D.

Following the general theorem, the next statements discuss the
specific forms of robust control Hamiltonian.

\textbf{Theorem:}\textit{\ (Necessary condition)}\textbf{\ }Controls in two non-commutative directions are necessary to correct the quasi-static noises coupled to all $x,y,z$ directions. That is $H_{c}(t)=\Omega_{j}(t)\sigma _{j}+\Omega _{k}(t)\sigma _{k}$, where $[\sigma _{j},\sigma_{k}]=i2\epsilon _{jkl}\sigma _{l}$.

For any given $j$ noise, $H_{c}$ should contain at least one term that is non-commutable with $\sigma _{j}$. Eq.~(\ref{Eq_XYdrive}) gives an example where $j=x$, $k=y$. So it is easy to demonstrate that for general $V=\boldsymbol{\delta }\cdot \hat{\boldsymbol{\sigma}}$, we get $[V,\Omega _{j}(t)\sigma _{j}+\Omega _{k}(t)\sigma_{k}]=i\boldsymbol{\gamma }\cdot \hat{\boldsymbol{\sigma }}$, where vector $\boldsymbol{\gamma }(t)\neq 0$ for $t\in \lbrack 0,T_{g}]$. 

Thus far, we have elucidated the necessary conditions for the implementation of robust control. This elucidation assists in clarifying the scenarios wherein dynamic error correction is infeasible. Moreover, these conditions could potentially be generalized to higher-dimensional systems. However, the feasibility of executing robust control to rectify arbitrary noisy quantum evolution has yet to be ascertained. Additionally, the existence of smooth and simplistic pulse sequences for robust control remains an open question. To address this, we hereby explore a conjecture. We will subsequently present both analytical and numerical solutions to the universal robust gates, which are capable of correcting errors induced by omnidirectional noise. These findings serve as a validation of the conjecture.

\textbf{Conjecture: }\textit{(Sufficient condition)}\textbf{\ }Controls in two directions are sufficient to correct the quasi-static noises coupled to all $x,y,z$ directions. That is $H_{c}(t)=\Omega _{j}(t)\sigma _{j}+\Omega_{k}(t)\sigma _{k}$, where $[\sigma _{j},\sigma _{k}]=2i\epsilon
_{jkl}\sigma _{l}$.

\section{Constructing robust control Hamiltonian}

\label{Sec_protocol}

In this section, we will show how to utilize the geometric correspondence of quantum evolution to quantify the robustness of arbitrary quantum control pulses and further present a protocol to construct the universal robust control Hamiltonian for arbitrary noises.

\subsection{Robustness measure}

\label{Sec_Measure}

\begin{figure}[t]
\centering
\includegraphics[width=0.8\columnwidth]{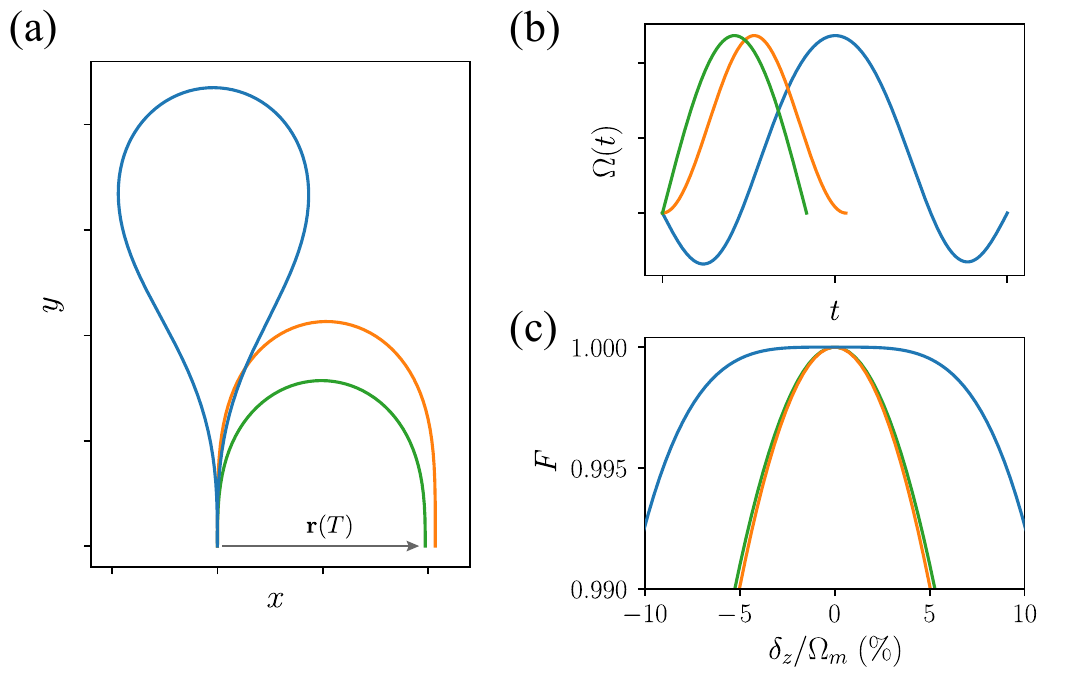}
\caption{(a) The QEED shows three error curves of RCP (blue), cosine (orange), and sine (green) pulses, corresponding to the pulse waveforms shown in (b). (c) Comparison of the fidelities as a function of the noise} strength for different pulse waveforms, as a demonstration of the agreement between the robustness behaviors of different pulses and their error distances. Here $\Omega_m=\max\{|\Omega(t)|\}$.
\label{Fig_RobustMeasure}
\end{figure}

The multiplex error curves, being the geometric correspondence of the error quantum dynamics, provide a geometric way to measure the error of any control pulses subject to specific noises. Specifically, let's define the \textit{error distance} to be the norm of the linear error term $|r(T_{g})|=||\mathbf{r}^{j}(T_{g} )||$, as in Eq.~(\ref{Eq_1order}). Unlike gate fidelity describing the overall performance of a quantum operation, the error distance $|r(T_{g})|$ characterizes the gate error subject to a specific noise-$j$. Therefore, quantitatively characterizing the robustness of a given pulse could be done by mapping the evolution to an error curve and then measuring the error distance between the starting point and ending point.  To show how the error distance characterizes the robustness of control pulses, we plot the z error curves of a robust pulse $R_{1;\bot}^{\pi }$ (defined in Sec.~\ref{Sec_UniversalSet}) and the two commonly used pulses in experiments, the cosine and sine pulses. As shown in Fig.~\ref{Fig_RobustMeasure} (a,b), the error distances associated with these three pulses exhibit different lengths. Note that the coordinates of the planar error curves in the manuscript are all denoted by $\{x,y\}$ for convenience. To verify the monotonicity of the robustness measure, we numerically simulate the driven noisy dynamics and obtain the gate fidelity of the three pulses as a function of the noise strength $\delta_z$ as shown in Fig.~\ref{Fig_RobustMeasure}(c). The fidelity of sine and cosine pulses cannot maintain a fidelity plateau as the RCP does, which agrees with their error distances illustrated in Fig.~\ref{Fig_RobustMeasure}(a). 

Therefore, the error distances are a good measure of robustness, and they will be employed as constraints within the pulse construction protocol delineated below. For a more quantitative examination of the robustness order across various pulses, the reader is referred to Appendix~\ref{AppendixB}. Due to the linear nature of error distances, cumulative errors arising from multiple noise sources are additive. This methodology for assessing noise susceptibility can be extended to quantum systems manifesting multiple error origins with multiplex error curves. Achieving simultaneous robustness across all error sources necessitates that the error distances for all curves vanish. Instances of such scenarios will be discussed in Section~\ref{Sec_UniversalSet}. Additionally, the second-order error $\Vert \mathbf{R}^{j}(T_{g} )\Vert $, as articulated in Eq.~(\ref{Eq_2order}), provides an extra dimension of higher-order control robustness, which will be expounded in Section~\ref{Sec_UniversalSet} and in Appendix~\ref{AppendixD}. It should be noted that a rigorous mathematical treatment of the geometric aspects of control robustness, grounded in measure theory, is outside the purview of this manuscript.

\subsection{Pulse construction protocol}

\label{Sec_PulseProtocol}

The geometric correspondence helps us understand how the noise affects the qubit dynamics. Therefore, it provides a way to find the conditions of robust evolution and motivates a pulse construction protocol by reverse engineering analytic space curves that satisfy the robustness conditions. The procedure of this construction is summarized as follows: (1) Construct a regular curve that meets certain boundary conditions determined by the target gate operation and robustness conditions such as closeness and vanishing net area; (2) Reparameterize the curve in terms of the arc-length parameter to make it moving at unit-speed; (3) Scale the length of the curve to fit an optional gate time; (4) Calculate its curvature and torsion to obtain the corresponding robust control pulses. This protocol is illustrated explicitly in previous works, where plane and space curves were used to construct different RCPs \cite{zeng2018general, zeng2019geometric,barnes2022dynamically}. We note that in our generic geometric correspondence, the control pulses are related to signed curvature and singularity-free torsion of the geometric curve, which are obtained by a set of well-chosen continuous Frenet vectors. We demonstrate the pulse construction
from space curves and provide a universal plane curve construction for first- and second-order robust pulses in the Appendix.~\ref{AppendixC} and~\ref{AppendixD}.

However, even with the constraints given by the theory, it is still challenging to make a good guess of ansatz for the RCPs. In a real quantum
computer, the universal gates using RCPs are desired to be generated automatically given the system parameters. Therefore, assistance with
numerical search is needed to find the RCPs automatically using the
robustness conditions.

Here, we present an analytic-numerical protocol to construct universal error-robust quantum gates that are made automatic. The theory of geometric correspondence gives analytical constraints for robustness, which is added to the objective function to perform constrained optimization using the COCOA algorithm \cite{song2022optimizing}. The protocol is described as follows.

\begin{figure}[htb]
\centering
\includegraphics[width=0.9\columnwidth]{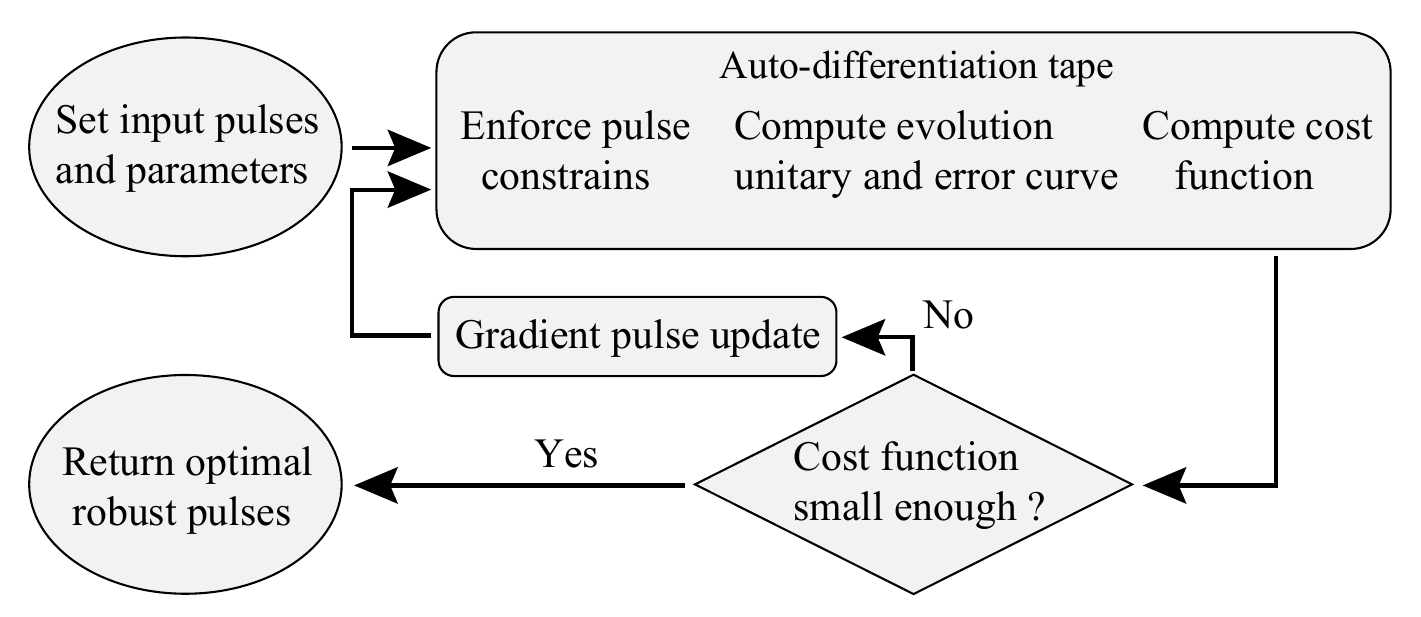}
\caption{ The flow chart of the pulse construction protocol. }
\label{Fig_flowchart}
\end{figure}

\begin{figure*}[hbt!]
\centering
\includegraphics[width=1\textwidth]{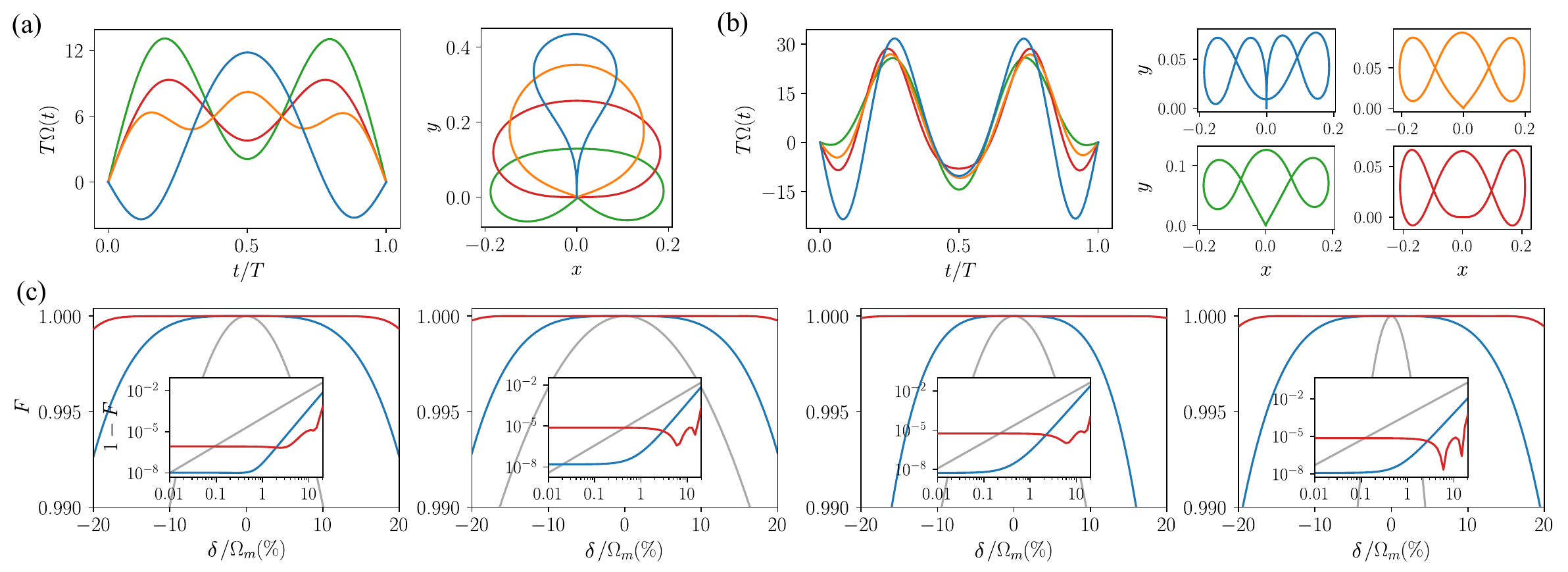}
\caption{(a) First order robust control pulses $\{ R_{1;\bot }^{\protect\pi},R_{1;\bot }^{7\protect\pi/4}, R_{1;\bot }^{5\protect\pi/2}, R_{1;\bot }^{2\protect\pi} \}$ (blue, orange, green, red) and their corresponding error curves. (b) Extended-robust control pulses $\{ R_{\text{ex};\bot}^{\protect\pi},R_{\text{ex};\bot}^{9\protect\pi/4},R_{\text{ex};\bot}^{5\protect\pi/2}, R_{\text{ex};\bot}^{2\protect\pi} \}$ (blue, orange, green, red) and their corresponding error curves. (c) Fidelities of the four single-qubit gates $\{ X_{\protect\pi}, X_{\protect\pi/4}, X_{\protect\pi/2}, X_{2\protect\pi} \}$ (left to right) realized by the first-order RCPs (blue) as in (a), extended RCPs (red) as in (b), and their cosine pulse counterpart (grey) against detuning noise. Insets: Infidelity as a function of relative noise strength in log-log scale.} 
\label{Fig_robust_orders}
\end{figure*}

(1) Initialize. Set initial input pulses and other relevant parameters.

(2) Analytic pulse constraints. We apply discrete Fourier series as the generic ansatz of the RCPs, so that we can control the smoothness of the pulses by truncating the Fourier basis while limiting the number of parameters. For pulse amplitude, the Fourier series is multiplied with a sine function to ensure zero boundary values, i.e., zero starting and ending of the pulse waveform. The ansatz of the pulse amplitude and phase takes the form 
\begin{equation}
\begin{aligned} \Omega_0(a_j,\phi_j;t) &= \sin( \frac{\pi t}{T} )(a_0 +
\sum_{j=1}^n a_j \cos( \frac{2\pi j}{T} t + \phi_j) ) \\
\Phi_0(b_j,\psi_j;t) &= b_0 + \sum_{j=1}^n b_j \cos(\frac{2\pi j}{T}t + \psi_j),
\end{aligned}  \label{Eq_Pansatz}
\end{equation}
where $n$ is the number of Fourier components, which will be set to be $[1,4]$ for different gates shown in this paper. $\{ a_j,\phi_j, b_j,\psi_j\} $'s are parameters to be optimized. For each input pulse, the corresponding Fourier series is obtained by a Fourier expansion and truncation, and then the expansion function for the amplitude is multiplied with the sine function to obtain a modified pulse. We choose the Fourier series ansatz because it contains few parameters, is convenient for experiments and it allows easy control over pulse smoothness, bandwidth and pre-distortion based on the transfer function.

(3) Use the modified pulses to compute the dynamics to obtain the evolution operator $U(t)$, as well as the error curve $\mathbf{r}(t)$.

(4) Compute the cost function, which takes the form 
\begin{equation}
C=(1-F) + D 
,  
\label{Eq_CostF}
\end{equation}%
where $F$ is the gate fidelity defined in Ref. \cite{pedersen2007fidelity} and $D$
are the robustness constraints. When only the first-order robustness is considered, 
$D =|r(T_{g} )| $, which is the error distance as defined in Sec.~\ref{Sec_Measure}. The robustness against errors in different axes can be achieved by including the error distances of different error curves in the cost function. The auto-differentiation tape records the calculations in (2)-(4).

(5) Make a gradient update of the pulse to minimize $C$.

(6) Go back to step (2) with the updated pulse as input if the cost function is larger than a criterion $\eta$, such as $10^{-5}$.

(7) If 
$C < \eta$, break the optimization cycle and obtain the optimal robust pulse, which takes the analytical form of Eq.~(\ref{Eq_Pansatz}).

\subsection{Universal set of RCP}

\label{Sec_UniversalSet}

By employing the pulse construction protocol delineated in the preceding subsection, we can derive RCPs that fulfill the robustness criteria stipulated by the geometric equivalence. As an illustrative instance, we employ this protocol to identify a suite of RCPs capable of implementing X-axis rotations that are robust to detuning noise. Subsequently, Eq.~(\ref{Eq_Hfull}) transforms into
\begin{equation}
H=\frac{1}{2}\Omega (t)\sigma_{x}+\frac{1}{2}(\Delta+\delta) \sigma _{z}.
\label{Eq_H1}
\end{equation}%
When the transverse control signal is on resonance with the qubit, $\Delta =0$. The first-order RCPs, referred to $\{R_{1;\bot}^{\pi },R_{1;\bot}^{7\pi /4},R_{1;\bot}^{5\pi /2},R_{1;\bot}^{2\pi }\}$, are obtained and shown in Fig.~\ref{Fig_robust_orders}(a). Here $R_{1;\bot}^{\theta }$ represents the first order RCP for a rotation of angle $\theta$ that can correct errors due to noises in its perpendicular directions. As an example, they can be used to implement an $X_\theta$ gate for Hamiltonian in Eq.~\ref{Eq_H1}, or a $Y_\theta$ gate by replacing the $\sigma_x$ with $\sigma_y$ control. We use $X_{\theta}$ and $Y_{\theta}$ to represent rotations of angle $\theta$ around the X-axis and Y-axis of the Bloch sphere respectively. The same RCP could be used to implement two-qubit gates as introduced in Sec.~\ref{Sec_application}. We denote the maximum absolute pulse amplitude as $\Omega_m=\text{max}\{|\Omega(t)|\}$.

Searching for higher-order robust pulses requires adding more terms associated with the higher-order robustness constraints, e.g., the net area of the error curve, to the constraint term $D$ in the cost function Eq.~(\ref{Eq_CostF}). However, we found that simply adding these constraints hinders the convergence of the algorithm at an unacceptable rate because the computation of these net areas involves complicated integrals over the error curve. We settle this issue by adding additional terms to the cost function $C$ with the gate fidelity and error distance at a non-vanishing $\Delta $. Specifically, we use $C=\sum_{j=1,2}(F(\Delta _{j})+\Vert r(\Delta _{j},T_{g})\Vert )$, where $\Delta _{1}=0$ and $\Delta _{2}$ are chosen to be $\pi/T_{g} $ according to the experimentalist's expectation. This ensures the resulting RCPs maintain high gate fidelities over a broader range of noise amplitude and thus lead to extended robustness. We then obtained a class of extended RCPs denoted as $\{R_{\text{ex};\bot}^{\pi },R_{\text{ex};\bot}^{9\pi /4},R_{\text{ex};\bot}^{5\pi /2}, R_{\text{ex};\bot}^{2\pi }\}$ that are made by three cosine functions and the error curves of them have net areas close to zero, as shown in Fig.~\ref{Fig_robust_orders}(b). The robustness of the eight RCPs mentioned above is further confirmed by calculating the first few Magnus expansion terms, as shown in the Appendix.~\ref{AppendixB}.

We use the first-order RCPs and the extended RCPs to implement four single-qubit robust gates $\{ X_{\pi}, X_{\pi/4}, X_{\pi/2}, X_{2\pi} \}$. 
The first three gates together with robust $Y_{\pi/2}$ gate, i.e., applying $R_{1;\bot}^{5\pi /2}$ or $R_{\text{ex};\bot}^{5\pi /2}$ in the Y direction, are elementary single-qubit gates for a universal robust gate set \cite{shor199637th}, while the $2\pi$ gate is equivalent to the robust dynamical decoupling. We numerically simulate the dynamics of the noisy qubit driven transversely (Eq.~\ref{Eq_H1}) and present the gate robustness in Fig.~\ref{Fig_robust_orders}(c). Compared with the gates performed by trivial cosine pulses with the same maximal amplitudes of the RCPs, the gates using the first-order RCPs and extended RCPs all exhibit wide high-fidelity plateaus over a significant range of noise amplitude. The gate infidelities exhibit plateaus around $10^{-8}$ within $1\%$ noise region for first-order RCPs and $10^{-5}$ within $15\%$ noise region for extended RCPs. Note that high-fidelity plateaus are a sign of the robustness of the gates.


\begin{figure}[t]
\centering
\includegraphics[width=1\columnwidth]{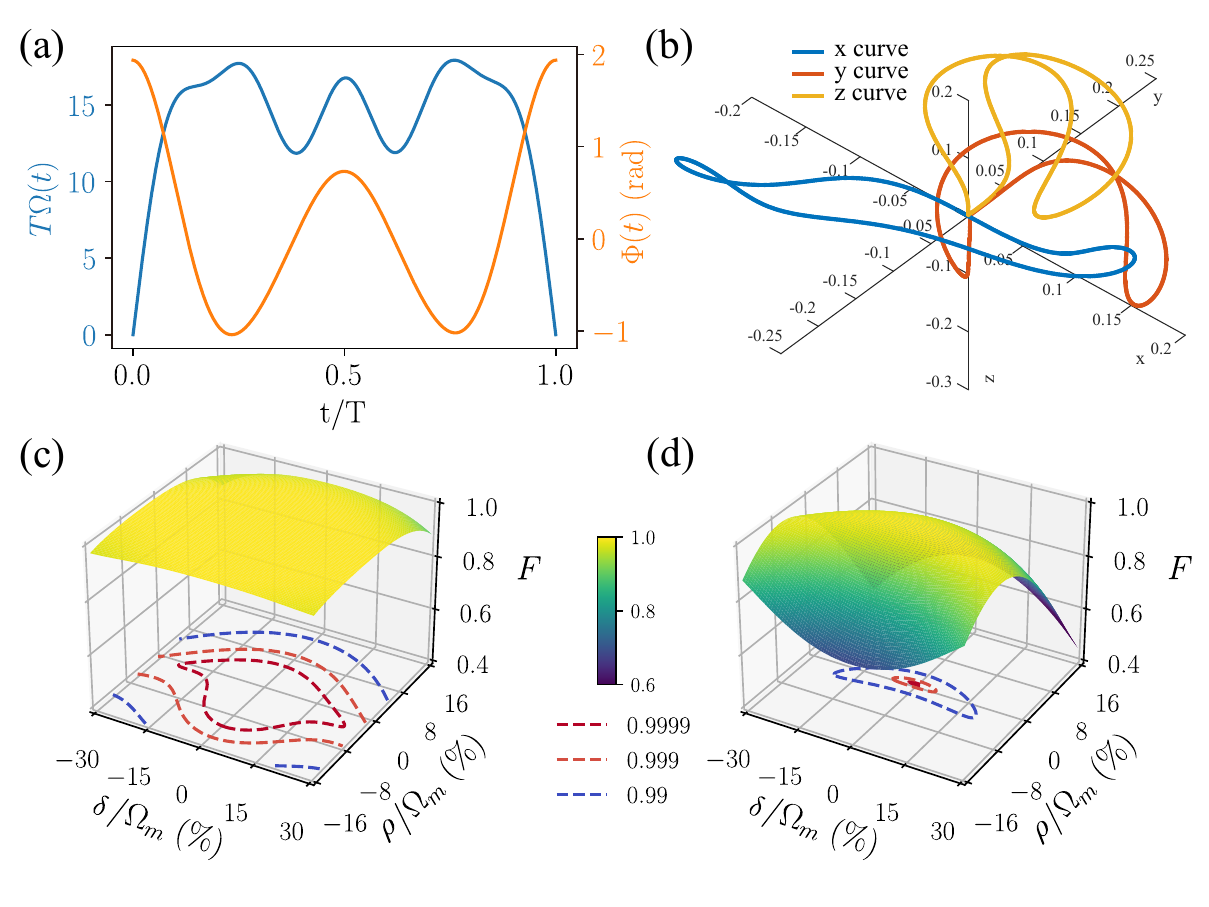}
\caption{ (a) A set of RCPs $R_{1;\text{all}}^{3\protect\pi/2}\{\Omega(t),\Phi(t)\}$
are used to implement robust gate $X_{3\protect\pi/2}$ to correct all errors in x, y, z directions, with the closed x, y, z error curves (blue, orange, yellow) shown in the QEED (b). A comparison of gate fidelity against longitudinal and transverse noises $\protect\delta$ and $\protect\rho$} using this $R_{1;\text{all}}^{3\protect\pi/2}$ and cosine pulse are shown in (c) and (d).
\label{Fig_XYdrive}
\end{figure}

More general noises on all axes can be corrected simultaneously with XY control with the Hamiltonian in Eq.~(\ref{Eq_XYdrive}). This works because the robust control in a direction could correct the noise coupled to the other two perpendicular directions, as discussed in the previous section. As a numerical demonstration, we consider both longitudinal and transverse noises in the form of $V=\frac{\delta }{2}\sigma_{z}+\frac{\rho }{2}(\sigma_{x}+\sigma_{y})$, giving rise to the co-existence of x, y and z error curves in the geometric space. Then the robustness constraint in Eq.~(\ref{Eq_CostF}) takes the form $D = \sum_{j=x,y,z}\Vert \mathbf{r}^{j}(T_{g} )\Vert $ We apply our protocol to find an RCP $R_{1;\text{all}}^{3\pi /2}$ with only four Fourier components to implement a single-qubit $X_{3\pi /2}$ gate that is robust against errors in all three directions. Although this RCP is solved to generate a robust $3\pi/2$ rotation around the x-axis, its rotation axis can be changed by adding a constant phase while keeping the robustness along the rotation axis and its two perpendicular axes, as demonstrated numerically in the Appendix.~\ref{AppendixB}. As shown in Fig.~\ref{Fig_XYdrive}(a) the XY drive has three closed error curves during the gate time. The fidelity landscape of the $X_{3\pi /2}$ gate using the $R_{1;\text{all}}^{3\pi /2}$ and the cosine pulse in the two error dimensions is plotted in Fig.~\ref{Fig_XYdrive}(b). The RCP shows a great advantage over the cosine pulse, showing a significantly high-fidelity plateau with fidelity above $0.9999$.

\section{universal robust quantum gates for realistic qubits}

\label{Sec_application}

Applying this method to construct universal robust quantum gates for realistic systems, especially in a multi-qubit setup, is not a trivial task. In this section, we demonstrate the method by studying the physical model of gate-defined quantum dots and superconducting transmon qubits.

Without loss of generality, the simulated gate time for all the RCPs is set to be $T_{g} =50$ ns. The realistic gate time can be arbitrary, and we can always rescale the RCPs in the time domain and maintain their robustness. This is guaranteed by the geometric correspondence since the substitution of $t\rightarrow \alpha t$, $\Omega \rightarrow \Omega /\alpha $ only rescales the length of the error curve and does not change the correspondence such as Eq.~(\ref{Eq_z_correspondence2}), as well as the unit-speed properties and robustness conditions of the error curves. In Appendix.~\ref{AppendixB}, we demonstrate this property and provide the parameters for the RCPs involved in the manuscript.


\subsection{Gate-defined quantum dot qubit}

\begin{figure}[htb]
\centering
\includegraphics[width=1\columnwidth]{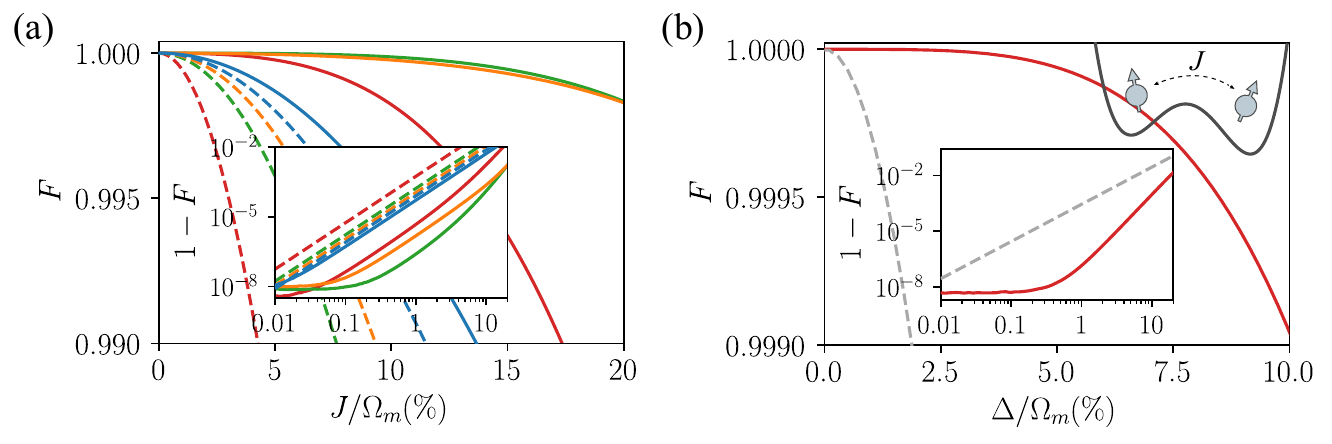}
\caption{ (a) Fidelities of single-qubit gate $\{ X_{\protect\pi},X_{\protect%
\pi/4},X_{\protect\pi/2},\newline
X_{2\protect\pi} \}$ (blue, orange, green, red) by RCPs (solid lines) and their cosine counterparts (dashed lines) as a function of the relative strength of the unwanted coupling $J$. (b) Fidelities of $\protect\sqrt{\text{SWAP}}$ gate by $R_{1;\bot }^{5\protect\pi/2}$ pulse (red) and cosine pulse (grey) as a function of the Zeeman difference. }
\label{Fig_QD}
\end{figure}

First, we consider a gate-defined double quantum dot (QD) system, in which the spin states of electrons serve as a qubit \cite{xue2022quantum, noiri2022fast,mkadzik2022precision, philips2022universal, huang2019fidelity}. The Hamiltonian is
\begin{equation}
H=\mathbf{B}_{1}\cdot \mathbf{S}_{1}+\mathbf{B}_{2}\cdot 
\mathbf{S}_{2}+J(\mathbf{S}_{1}\cdot \mathbf{S}_{2}-1/4),
\label{Eq_HeisenbergEq}
\end{equation}%
where $\mathbf{S}_{j}=(\sigma_{x,j},\sigma_{y,j},%
\sigma_{z,j})/2$ and $\mathbf{B}_{j}=(B_{x,j},B_{y,j},B_{z,j})$ are the spin operators and magnetic field acting on qubit $j$, $J$ is the exchange interaction between the spins. We denote the difference of Zeeman splitting between the two qubits as $\Delta=B_{z,2}-B_{z,1}$. Detuning noise from magnetic fluctuations, spin-orbit interaction, and residual exchange interaction are the major obstacles to further improving the coherence and gate fidelity of such qubits \cite{xue2022quantum, bertrand2015quantum}.

\textit{Single-qubit robust gates.}-A single QD is a good two-level system. With detuning noise coupled to a driven QD, the system Hamiltonian takes the same form as Eq.~(\ref{Eq_H1}). Therefore, the robustness of the pulses driving a single QD can be well described by Fig.~\ref{Fig_robust_orders}.

Another source of single-qubit errors is the unwanted coupling between two QDs. The Hamiltonian of such a system with Heisenberg interaction takes the form of Eq.~(\ref{Eq_HeisenbergEq}). Ideally, the universal single-qubit gates $\{ X_{\pi}, X_{\pi/4}, X_{\pi/2}, X_{2\pi} \}$ on the second qubit are given by $I\otimes X_{\theta}$, as the operations on the second qubit shall not affect the first qubit. It is reasonable to assume $J \ll \Delta$, but $J$ cannot be completely turned off during the single-qubit operations by tuning the gate voltage between two QDs \cite{xue2022quantum}. In the eigenbasis, the two subspaces $\text{span}\{ | \tilde{\uparrow \uparrow} \rangle, | \tilde{\uparrow\downarrow} \rangle \}$ and $\text{span}\{ | \tilde{\downarrow \uparrow}\rangle, |\tilde{\downarrow \downarrow} \rangle \}$ both correspond to the second qubit but are detuned by $J$ in the rotating frame. We tune a magnetic drive $B_{x,2}(t)$ according to the first order RCPs to implement robust $X$ rotations against this detuning resulting from the unwanted coupling. The numerical results demonstrating robust $\{X_{\pi},X_{\pi/4},X_{\pi/2},X_{2\pi} \}$ gates in such double QD system with $\Delta = 250$ MHz are shown in Fig.~\ref{Fig_QD}(a). Although what we demonstrated here is a partial error cancellation since there is an additional crosstalk error between the two subspaces, which will reduce the gate robustness to some extent, the fidelities are still significantly improved in comparison with the cosine waveform. Further, the crosstalk error can be canceled by appropriate timing and by implementing additional correction pulses according to \cite{russ2018high, huang2019fidelity}, or with more sophisticated RCPs. 

\textit{Two-qubit robust gates.}-Two-qubit $\sqrt{\text{SWAP}}$ gate is commonly used as the entangling gates in the QD system. A perfect $\sqrt{\text{SWAP}}$ can be realized by switching on the exchange coupling in the region where two QDs have zero Zeeman difference---a condition hardly met in experiments either due to the detuning noise or the need for qubit addressability. Note that the entangling originates from the Heisenberg interaction and the Hamiltonian Eq.~(\ref{Eq_HeisenbergEq}) in the anti-parallel spin states subspace $\text{span}\{ \left|\uparrow \downarrow \right\rangle, \left|\downarrow \uparrow \right\rangle \}$ takes the form 
\begin{equation}
H(t) = \frac{1}{2} \left(%
\begin{array}{cc}
-J(t) + \Delta & J(t) \\ 
J(t) & -J(t) - \Delta%
\end{array}%
\right).
\end{equation}
To implement a $\sqrt{\text{SWAP}}$ gate means to control $J(t)$ to
implement $X_{\pi/2}$ operation in this subspace. The waveform of $J(t)$ could be upgraded to a robust pulse $R_{1;\bot }^{5\pi/2}$ in order to fight against the finite Zeeman difference. We numerically solve the time-dependent Schr\"odinger equation and obtain the results shown in Fig.~\ref{Fig_QD}. The robust $\sqrt{\text{SWAP}}$ gate exhibits a high-fidelity plateau, and the infidelity is several orders of magnitude lower than the cosine pulse counterpart in a wide detuning region.


\subsection{Transmon qubit}

\textit{Single-qubit gates robust to frequency variation.}-For a single superconducting transmon with qubit frequency variation, we use the RCPs $\{R_{1}^{\pi },R_{1}^{7\pi /4},R_{1}^{5\pi /2},R_{1}^{2\pi }\}$ found in Fig.~\ref{Fig_robust_orders} to implement robust single-qubit gates $\{X_{\pi },X_{\pi /4},X_{\pi /2},X_{2\pi }\}$. A transmon is usually considered as a three-level system with moderate anharmonicity between the first and second transitions \cite{blais2021circuit, krantz2019quantum}. With the transverse control field, a transmon could be modeled as  \begin{equation}
H=(\omega +\delta )a^{\dagger }a+\frac{u}{2}a^{\dagger }a^{\dagger
}aa+H_{c}(t).
\end{equation}%
where the control Hamiltonian \cite{krantz2019quantum} 
\begin{equation}
H_{c}(t)=\frac{1}{2}\xi (t)ae^{i\omega _{d}t}-h.c.
\end{equation}%
The lowest two levels form the qubit. Here $\omega $, $u$, $a^{\dagger }$, $a$ are the qubit frequency, the anharmonicity, and the creation and annihilation operators of the transmon. $\omega _{d}$ is the frequency of the driving and $\xi (t)$ is the total control waveform. We apply our theory to the dynamics associated with the qubit levels to correct the errors resulting from frequency variation, and use DRAG to suppress the leakage to higher levels \cite{motzoi2009simple}. So $\xi (t)$ is a sum of RCP $\Omega (t)$ and the corresponding DRAG pulse $i\alpha \dot{\Omega}(t)$. The DRAG parameter $\alpha $ is related to the transmon anharmonicity \cite{motzoi2009simple} and takes the numerically optimized value at zero detuning in our simulation.

\begin{figure}[t]
\centering
\includegraphics[width=1\columnwidth]{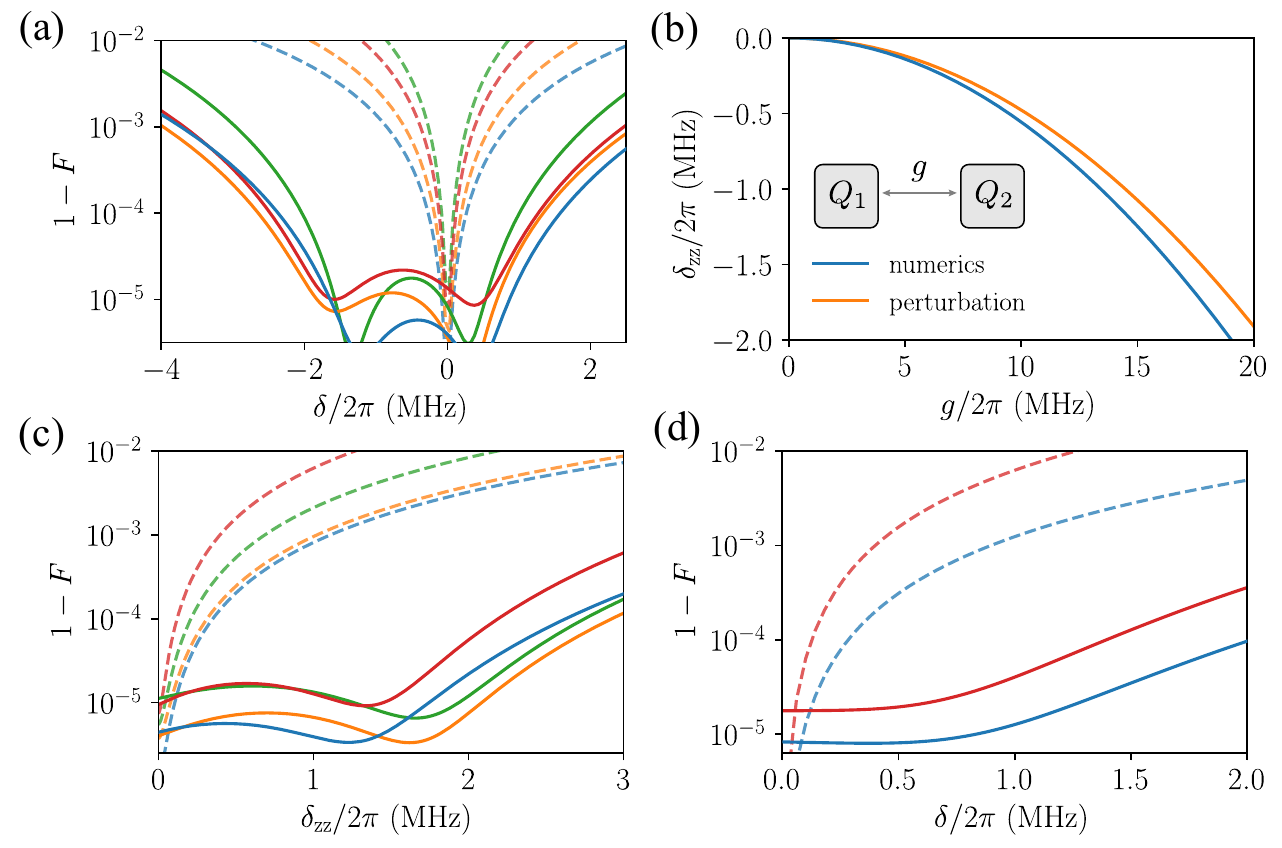}
\caption{ (a) Fidelities of single-qubit gates $\{ X_{\protect\pi},X_{\protect\pi/4},X_{\protect\pi/2},\newline
X_{2\protect\pi} \}$ (blue, orange, green, red) for a transmon qubit by RCPs (solid) and their cosine counterparts (dashed) as a function of qubit frequency variation. (b) The strength of ZZ-coupling $\protect\delta_{zz}$ for two capacitively interacting transmons solved numerically (blue) and analytically using perturbation theory (orange). $g$ is the unwanted interaction strength. (c) Fidelities of single-qubit gates $\{ X_{\protect\pi}, X_{\protect\pi/4}, X_{\protect\pi/2}, X_{2\protect\pi} \}$ (blue, orange, green, red) for $Q_2$ by robust pulses (solid) and their cosine counterparts (dashed) as a function of residual ZZ-coupling strength caused by unwanted interaction $g$. (d) Fidelity of two-qubit iSWAP (blue) and $\protect\sqrt{\text{iSWAP}}$ (red) for two coupled transmon qubits by RCPs (solid) and their cosine counterparts (dashed) as a function of the qubits' frequency difference. }
\label{Fig_transmon}
\end{figure}

We set the gate times for $\{X_{\pi },X_{\pi /4},X_{\pi /2},X_{2\pi }\}$ to be $70,50,80$ and $55$ ns such that the maximal pulse amplitudes $\Omega_{m}/2\pi $ are around $27$ MHz, and take the anharmonicity $u/2\pi = -0.26$ GHz. Like previous sections, we use cosine pulses with the same maximal amplitudes for comparison. Our numerical results in Fig.~\ref{Fig_transmon}(a) illustrate the fidelity plateau over a few MHz of frequency variation for our robust gates. Note that the centers of the fidelity plateau all shift to the left. We conclude the reason to be the AC-Stark shifts associated with the higher levels.

\textit{Single-qubit gates robust to unwanted coupling.-}In realistic multi-qubit systems, a transmon is unavoidably coupled to another quantum system, the so-called spectators \cite{deng2021correcting,zhao2022quantum,krinner2020benchmarking, wei2022hamiltonian,kandala2021demonstration,mundada2019suppression, ku2020suppression}. This results in qubit frequency splittings or spectrum broadening, giving rise to correlated errors, which become a major obstacle for large-scale quantum computing \cite{krinner2020benchmarking}. As a demonstrative example, we consider two directly coupled transmon qubits (one spectator and one target qubit) with the Hamiltonian 
\begin{equation}
H_{0}=\sum_{j=1,2}[\omega _{j}a_{j}^{\dagger }a_{j}+\frac{u_{j}}{2}%
a_{j}^{\dagger }a_{j}^{\dagger }a_{j}a_{j}]+g(a_{1}^{\dagger
}a_{2}+a_{1}a_{2}^{\dagger }),  \label{Eq_2Transmon}
\end{equation}%
where $\omega _{j}$ is the qubit frequency, $u_{j}$ is the anharmonicity for the $j$-th qubit and $g$ is the interaction strength. Denote the eigenstates of the spectator-target qubit system \cite{deng2021correcting} as $|S,T\rangle $. Up to the second-order perturbation, the effective diagonal Hamiltonian in the subspace $\{\left\vert 00\right\rangle ,\left\vert01\right\rangle \}$ and $\{\left\vert 10\right\rangle ,\left\vert11\right\rangle \}$ is detuned by $\delta _{\text{zz}}$ with $\delta _{\text{zz}}=-\frac{2g^{2}(u_{1}+u_{2})}{(u_{1}+\Delta )(u_{2}-\Delta )}$ and $\Delta =\omega _{2}-\omega _{1}$. This detuning $\delta _{\text{zz}}$ is known as the residual ZZ-coupling. We take $\omega _{1}/2\pi =5.0$ GHz, $\omega _{2}/2\pi =5.5$ GHz, $u_{1}/2\pi =-0.23$ GHz, $u_{2}/2\pi =-0.26$ GHz. The exact $\delta _{\text{zz}}$ obtained by numerical diagonalization and the perturbative one are plotted in Fig.~\ref{Fig_transmon}(b) as a demonstration of the effectiveness for the perturbative treatment of $\delta_{\text{zz}}$ at small $g$. Our numerical results as shown in Fig.~\ref{Fig_transmon}(c) illustrate a significant improvement of gate infidelity in the presence of residual ZZ-coupling up to the order of MHz. When ZZ-coupling is vanishing, the infidelity of robust gates remains of the same order as trivial pulses. Here the DRAG control \cite{motzoi2009simple} is added as well.

\textit{Two-qubit gates.-}iSWAP gate is one of the favorable two-qubit entangling gates for quantum computing using transmons \cite{arute2019quantum}. Traditional implementation of the iSWAP gate requires the two qubits to be on resonance. This could be achieved by tuning qubit frequencies, which, however, induces extra flux noise and hence the variation in qubit frequency. To better tune the interaction strength, recent architecture favors tunable couplers \cite{xu2020high, stehlik2021tunable}. This doesn't change the effective Hamiltonian in Eq.\ref{Eq_2Transmon}, only enabling the control of the coupling strength $g(t)$. To correct the error due to the uncertain frequency difference $\delta $ between the two transmons, we apply RCPs $\{R_{1}^{\pi },R_{1}^{5\pi /2}\}$ on $g(t)$ to implement the robust iSWAP and $\sqrt{\text{iSWAP}}$ gates. We use the qubit settings as in the previous discussion to study the variation of gate infidelities as a function of $\delta $. The additional ZZ-coupling induced by higher levels of the transmons when activating the iSWAP interaction is omitted in our simulation as it can be canceled by exploiting the tunable coupler and setting a proper gate time \cite{sung2021realization}. Our numerical results in Fig.\ref{Fig_transmon}(d) demonstrate a great advantage of RCPs over their cosine counterparts. This control scheme also illustrates the great potential to simplify the experimental tune-ups of qubit frequencies. 

\section{Conclusion and discussion}
\label{Sec_Conclude}

In summary, the geometric correspondence generates the quantum error evolution diagrams as a graphical analysis for studying the error dynamics of driven quantum systems subject to generic noises. The geometric measures of the errors obtained from the diagrams (Sec.~\ref{Sec_Measure}) enable the intuitive and quantitative description of the robustness of arbitrary quantum operations. Furthermore, the framework provides conditions (Sec.~\ref{subsecRobCon}) to identify feasible robust control scenarios and allows us to establish an experimental-friendly robust control protocol (Sec.~\ref{Sec_PulseProtocol}) as a fusion of analytical theory and numerical optimization. Here, this protocol exploits no more than four cosine components to construct universal robust quantum gates that can tackle generic errors. These simple pulses will enable low-cost generation and characterization in experiments. The duration of the generated pulses is flexibly adjustable while maintaining robustness, as shown in Fig.\ref{Fig_pulse_rescale}. The applications to realistic systems are demonstrated by the numerical simulations on semiconductor spin and superconducting transmon systems, where the gate performance is shown beyond the fault-tolerance threshold over a broad robust plateau. With this simple and versatile protocol, our robust control framework could be applied to experiments immediately and adopted on various physical systems to enhance their robustness in complex noisy environments. The data of the robust control pulses presented in the manuscript and a demonstration of our pulse construction algorithm are available on GitHub \cite{GitHubRepo}. 

The geometric correspondence is essential. Although our discussion in this manuscript focuses on single- and two-qubit systems with multiple quasi-static noises, the framework is directly extensible to multi-qubit systems with complex, time-dependent noise via higher-dimensional geometric correspondence and error curves with modified speeds. Hence it shows a promising prospect in large quantum processors, and will potentially inspire a wave of study on noisy quantum evolution, quantum control and compiling, error mitigation and correction in quantum computers, quantum sensing, metrology, etc. On the other hand, a rigorous mathematical study of the geometric measure theory remains open and is beyond the scope of this work.

\acknowledgments
XHD conceived and oversaw the project. YJH and XHD derived the theory, designed the protocols, and wrote the manuscript. YJH did all the coding and numerical simulations. YS, JL and JZ gave some important suggestions on the algorithm. All authors contributed to the discussions.

We thank Yu He, Fei Yan for suggestions on the simulations of the realistic models and Qihao Guo, Yuanzhen Chen for fruitful discussions. This work was supported by the Key-Area Research and Development Program of Guang-Dong Province (Grant No. 2018B030326001), the National Natural Science Foundation of China (U1801661), the Guangdong Innovative and Entrepreneurial Research Team Program (2016ZT06D348), the Natural Science Foundation of Guangdong Province (2017B030308003), and Shenzhen Science and Technology Program (KQTD20200820113010023).

\begin{appendix}
\section{Geometric Correspondence}
\label{AppendixA}
Without loss of generality, we first derive the geometric correspondence given by the z-error curve discussed in the main text. The Hamiltonian in the interaction picture and the error curve is
\begin{eqnarray}
H_{I}(t) &=& \delta U_{0}^{\dagger }(t)\sigma_{z}U_{0}(t)= \delta \mathbf{T} \cdot \hat{\boldsymbol{\sigma}} \\
\mathbf{r}(t) \cdot \hat{\boldsymbol{\sigma}} &=& 
\int_{0}^{t}U_{0}^{\dagger }(t_{1})\sigma_{z}U_{0}(t_{1})dt_{1}.  \notag   \label{Eq_A_tangent}
\end{eqnarray}%
This noise term $\delta$ is treated as a static perturbation, which agrees with the physical
picture where the duration of a quantum gate is much shorter than the time
the scale of most of the pink noise or decoherence. Note $\dot{\mathbf{r}}(t)=\mathbf{T}$ is a unit tangent vector, one can then
obtain a new unit vector $\mathbf{N}$ perpendicular to $\mathbf{T}$ through 
\begin{eqnarray}
\dot{\mathbf{T}}(t)\cdot\hat{\sigma} &=&iU_{0}^{\dagger
}(t)[H_{0}(t),\sigma _{z}]U_{0}(t)  \notag \\
&=&\Omega (t)U_{0}^{\dagger }(t)(-\sin \Phi (t)\sigma _{x}+\cos \Phi
(t)\sigma _{y})U_{0}(t)  \notag \\
&=&\Omega (t)\mathbf{N}\cdot \hat{\boldsymbol{\sigma}}.  \label{Eq_A_normal}
\end{eqnarray}%
The third vector $\mathbf{B}$ perpendicular to $\mathbf{T}$ and $\mathbf{N}$
is given by $\mathbf{B}=\mathbf{T}\times \mathbf{N}$, 
\begin{equation}
\mathbf{B}(t)\cdot \hat{\boldsymbol{\sigma}}=U_{0}^{\dagger }(t)(-\cos \Phi (t)\sigma
_{x}-\sin \Phi (t)\sigma _{y})U_{0}(t).
\label{Eq_A_binormal}
\end{equation}%
Its time derivative satisfies 
\begin{eqnarray}
\dot{\mathbf{B}}(t)\cdot \hat{\boldsymbol{\sigma}} &=&\dot{\Phi}U_{0}^{\dagger }(t)(\sin
\Phi (t)\sigma _{x}-\cos \Phi (t)\sigma _{y})U_{0}(t)  \notag \\
&=&-\dot{\Phi}(t)\mathbf{N}\cdot \hat{\boldsymbol{\sigma}}. \label{Eq_A_Dbinormal}
\end{eqnarray}

The three unit vectors $\{\mathbf{T},\mathbf{N},\mathbf{B}\}$, as tangent, normal, and binormal unit vectors of the error curve formed a Frenet-Serret frame, and their defining formulas Eq.~(\ref{Eq_z_correspondence1}) directly follows Eq.~(\ref{Eq_A_tangent})-(\ref{Eq_A_binormal}). They satisfy the Frenet-Serret equations 
\begin{equation}
\left( 
\begin{array}{l}
\dot{\mathbf{T}} \\ 
\dot{\mathbf{N}} \\ 
\dot{\mathbf{B}}%
\end{array}%
\right) =\left( 
\begin{array}{ccc}
0 & \kappa  & 0 \\ 
-\kappa  & 0 & \tau  \\ 
0 & -\tau  & 0%
\end{array}%
\right) \left( 
\begin{array}{l}
\mathbf{T} \\ 
\mathbf{N} \\ 
\mathbf{B}%
\end{array}%
\right),
\label{Eq_A_Frenet}
\end{equation}%
with $\Omega(t)$ and $\dot{\Phi}(t)$ play the role of signed curvature $\kappa(t)$ and singularity free torsion $\tau(t)$ of the error curve, as stated in Eq.~(\ref{Eq_z_correspondence2}). 


Different from standard differential geometry, the Frenet vectors defined physically by Eq.~(\ref{Eq_z_correspondence1}) are continuous and differentiable since the pulses are assumed to be continuous and differentiable. We call the signed curvature as the projection of the curvature vector $\dot{\mathbf{T}}$ onto the normal vector $\dot{\mathbf{T}}\cdot \mathbf{N}=\kappa$. It can take negative values since the corresponding pulse amplitude $\Omega$ can be negative. It relates to the conventional curvature in standard differential geometry of curves by taking the absolute value. The torsion defined by $\dot{\mathbf{B}} \cdot \mathbf{N}=-\tau $ is also continuous and does not have a singularity at curvature zero point as in standard differential geometry. This mathematical ambiguity is addressed in Section.~\ref{AppendixC}.

One can also establish the geometric correspondence by error curve in other directions. Here we take the x error curve as an example. The x- error curve is given by $\dot{\mathbf{r}} = U_0^{\dagger} \sigma_x U_0$, and the Frenet vectors are defined as
\begin{equation}
\begin{array}{c}
\mathbf{T}\cdot \hat{\boldsymbol{\sigma}} = U_{0}^{\dagger }(t)\sigma_{x}U_{0}(t) \\ 
\mathbf{N}\cdot \hat{\boldsymbol{\sigma}} = - U_{0}^{\dagger }(t)\sigma_{z}U_{0}(t) \\ 
\mathbf{B}\cdot \hat{\boldsymbol{\sigma}} = U_{0}^{\dagger }(t)\sigma_{y}U_{0}(t).
\end{array}%
\end{equation}
The relation between control pulses and the curvature-torsion of the x-error curve is given by
\begin{equation}
\begin{aligned}
\kappa(t)  &= \dot{\mathbf{T}} \cdot \mathbf{N} =  \Omega(t) \sin\Phi(t)   \\
\tau(t)  &= -\dot{\mathbf{B}}\cdot \mathbf{N} = \Omega(t) \cos\Phi(t).
\end{aligned}
\end{equation}
Imposing robustness constraints on this error curve will lead to the dynamics robust against x error.

\section{From regular space curve to pulse}
\label{AppendixC}

As mentioned in the main text, one can construct robust control pulses by reverse engineering of space curves because of the geometric correspondence between regular space curves and the dynamics of a two-level system, where the signed curvature and singularity-free torsion of the space curves are related to the drive amplitude and phase. However, from the mathematical point of view, it is known that the existence and continuity of the Frenet-Serret frame are not guaranteed in standard differential geometry in the vicinity of curvature-vanishing points since the definition of $\mathbf{N}, \mathbf{B}$ and torsion need the curvature to be non-zero \cite{o2006elementary, pressley2010elementary}. This singularity and discontinuity issue can be solved in a purely mathematical manner by defining the continuous Frenet-Serret frame vectors in terms of three singularity-free Frenet-Euler angles, as introduced in \cite{shabana2022curvature}. 
The corresponding control pulses of the dynamics can be obtained by the resulting signed curvature and singularity-free torsion in terms of an arc-length variable. We summarize this approach and demonstrate a few examples as follows.

For a regular space curve in arbitrary parametrization $\mathbf{r}(\lambda)$. The unit tangent vector $\mathbf{T}$ can always be written in terms of two angles by 
\begin{equation}
\mathbf{T} = (x^{\prime},y^{\prime},z^{\prime})^T / |\mathbf{r}^{\prime}| = (\cos\psi \cos\theta, \sin\psi \cos\theta, \sin\theta)^T,
\label{Eq_C_tangent}
\end{equation}
where $^\prime$ represents derivative with respect to $\lambda$ and angles $\psi$ and $\theta$ are determined by derivatives of the curve coordinate.

The existence of well-defined normal vectors relies on the third angle $\phi$ defined as
\begin{equation}
\tan \phi = -\frac{(z^{\prime\prime}(x^{\prime 2}+y^{\prime 2})-z^{\prime}(x^{\prime} x^{\prime\prime}+y^{\prime} y^{\prime\prime}))}{|\mathbf{r}^{\prime}| (y^{\prime\prime} x^{\prime} - x^{\prime\prime} y^{\prime})}.
\label{Eq_C_phi}
\end{equation}
At the curvature-zero point, where $\kappa=x^{\prime\prime}=y^{\prime\prime}=z^{\prime\prime}=0$, it is still well-defined by taking limit according to L'Hospital rule.

The other two vectors of the coordinate system are then expressed as
\begin{equation}
\begin{aligned}
\mathbf{N} = \left( \begin{array}{c}
 -\sin \psi \cos \phi+\cos \psi \sin \theta \sin \phi  \\
 \cos \psi \cos \phi+\sin \psi \sin \theta \sin \phi  \\
 -\cos \theta \sin \phi 
\end{array}\right)
\end{aligned}
\label{Eq_C_normal}
\end{equation}
and
\begin{equation}
\begin{aligned}
\mathbf{B} = \left( \begin{array}{c}
-\sin \psi \sin \phi-\cos \psi \sin \theta \cos \phi \\
\cos \psi \sin \phi-\sin \psi \sin \theta \cos \phi \\
\cos \theta \cos \phi
\end{array}\right).
\end{aligned}
\label{Eq_C_binormal}
\end{equation}
Here the three angles $\{\psi,\theta,\phi\}$, named Frenet angles, are used to define curve geometry and resolve the ambiguity at curvature-vanishing points. We refer to \cite{shabana2022curvature} for a detailed discussion of this choice of coordinate system. According to the Frenet equation, the signed curvature and singularity-free torsion are obtained from the continuous Frenet vectors by
\begin{equation}
\begin{aligned}
\kappa(\lambda) &= \frac{\mathbf{T}^{\prime}\cdot \mathbf{N}}{||\mathbf{r}^{\prime}||} \\
\tau(\lambda) &= - \frac{\mathbf{B}^{\prime}\cdot \mathbf{N}}{||\mathbf{r}^{\prime}||}.
\end{aligned}
\label{Eq_C_cur_tor}
\end{equation}

The geometric correspondence with the z error curve given by Eq.~(\ref{Eq_z_correspondence1}) and the above-mentioned mathematically non-singular frame choice can be connected as follows. Consider a conventional parametrization of evolution unitary by
\begin{equation}
\begin{aligned}
&U_{0}(t)=\left(\begin{array}{cc}
u_{1}(t) & -u_{2}^{*}(t) \\
u_{2}(t) & u_{1}^{*}(t)
\end{array}\right) \\
&u_{1}(t)=e^{\frac{1}{2} i(\psi_1(t) + \phi_1(t))} \cos \left(\frac{\theta_1(t)}{2}\right) \\
&u_{2}(t)=-i e^{\frac{1}{2} i(\psi_1(t)-\phi_1(t))} \sin \left(\frac{\theta_1(t)}{2}\right).
\end{aligned}
\label{Eq_C_para_unitary}
\end{equation}
By equating the Frenet vectors defined by Eq.~(\ref{Eq_C_tangent})-(\ref{Eq_C_binormal}) and Eq.~(\ref{Eq_z_correspondence1}) with the unitary parametrization of Eq.~(\ref{Eq_C_para_unitary}) we have the following relations of angles
\begin{equation}
\begin{aligned}
\psi_1 &= \psi - \frac{\pi}{2}   \\
\theta_1 &= \frac{\pi}{2} - \theta   \\
\phi_1 + \Phi &= \frac{\pi}{2} - \phi.
\end{aligned}
\end{equation}

As indicated in \cite{zeng2019geometric}, the initial condition of the ideal evolution and the target gate unitary at the given gate time $T$ set the boundary conditions of these angles. 
Note that $U_0(0)=I$ gives the initial conditions $\theta_1(0)=0$ and $\psi_1(0) = - \phi_1(0)$. Since $H_0(t) = i\dot{U}_0(t)U_0^{\dagger}(t)$, we have
$$
( i \dot{\psi_1} \sin\theta_1 + \dot{\theta}_1 ) e^{-i \phi_1} = \Omega(t)e^{i\Phi}
$$
and $\Phi(0) = - \phi_1(0) = \psi_1(0)$. The final $\theta_1(T)$ and $\psi_1(T)$ correspond to the final time tangent vector $\dot{\mathbf{r}}(T)$ and $\phi_1(T)$ is related to the total torsion
$$
\phi_1(T)-\phi_1(0) = -\int_{0}^{T} \tau(t) d t - \arg[ i \dot{\psi_1} \sin\theta_1 + \dot{\theta}_1 ]|_{0} ^{T}.
$$

\begin{figure}[th]
	\centering
	\includegraphics[width=1\columnwidth]{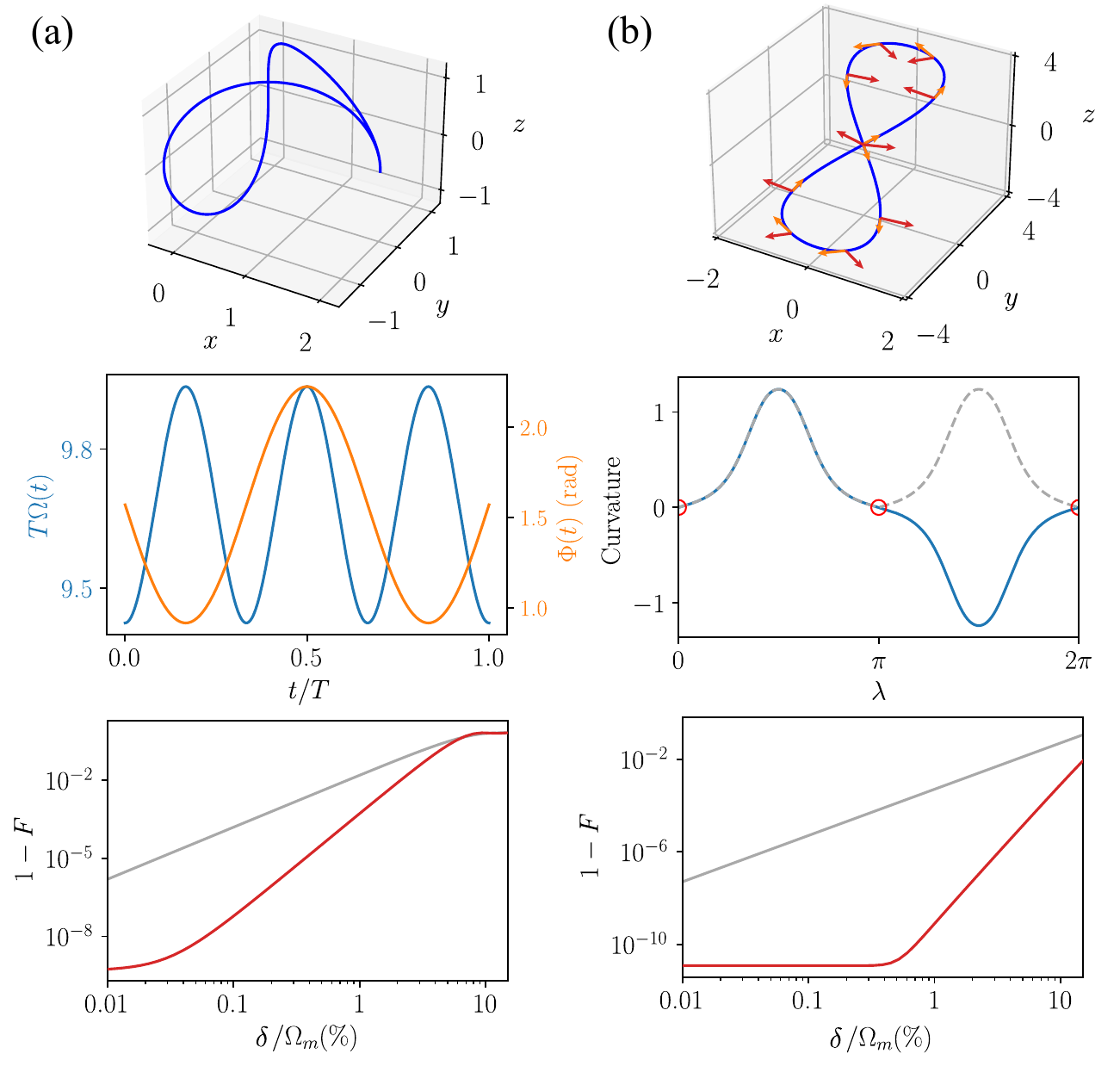}
	\caption{(a) A space curve in the QEED (up), the corresponding RCP (middle), and the fidelity of $X_{\pi}$ gate performed by the RCP (red) and its cosine pulse counterpart (grey) against detuning noise (down). (b) A figure 8 space curve with its tangent and continuous normal vectors marked by the orange and red arrows (up); the curvature (dashed grey) and signed curvature (blue) of this curve (middle). The curvature zero points are marked by red circles; The fidelity of $X_{2\pi}$ gate performed by the RCP obtained from this curve (red) and its cosine pulse counterpart (grey) against detuning noise (right).     }
	\label{Fig_spacecurve}
\end{figure}

Having established the correspondence from the space curve to the control pulse, we further elucidate this correspondence by constructing robust control pulses with the following examples of space curves. First, for a unit-speed space curve 
$$
\begin{aligned}
\mathbf{r}(t) = \left( \begin{array}{c}
(1+\cos \frac{t}{2} )\cos \frac{t}{2} \\
(1-\cos \frac{t}{2})\sin \frac{t}{2} \\
\frac{4}{3} \sin \frac{3t}{4}
\end{array}\right)
\end{aligned}
$$
with $t \in [0,4\pi]$. Note that in the context of curve geometry, we refer to $t$ as the curve length variable. One can determine the Frenet angles unambiguously from the tangent vector and Eq.~(\ref{Eq_C_tangent}) to obtain 
$\theta = - 3t/4 + \pi/2 $, $\psi = -t/4 + \pi$ and $\tan\phi = -3/\sin(3t/4)$. Then vectors $\mathbf{N}$ and $\mathbf{B}$ can be obtained straightforwardly, and the curvature and torsion are given by
$$
\begin{aligned}
\kappa(t) = \frac{1}{8} \sqrt{38 - 2 \cos \frac{3t}{2} } \\
\tau(t) = \frac{-73 \cos \frac{3t}{4}  + \cos \frac{9t}{4} }{-152 + 8\cos \frac{3t}{2}}.
\end{aligned}
$$

The corresponding pulses obtained from this curve via the correspondence Eq.~(\ref{Eq_z_correspondence2}) are shown in Fig.~\ref{Fig_spacecurve}(a). It generates a $\pi$ rotation if we choose an initial phase $\pi/2$ as the integral constant of $\Phi$. The robustness of $X_{\pi}$ gate performed by them against detuning noise are also shown in Fig.~\ref{Fig_spacecurve}(a).

Specifically, for plane curve, we have $\theta = \phi = 0$, $\mathbf{T}=(\cos\psi,\sin\psi,0)^T$ and $\mathbf{N}=(-\sin\psi,\cos\psi,0)^T$ with $\cos\psi = x^{\prime}/\sqrt{x^{\prime 2}+y^{\prime 2}}$ and $\sin\psi = y^{\prime}/\sqrt{x^{\prime 2}+y^{\prime 2}}$. In this case, the normal vector is well-defined regardless of the existence of curvature zero.  According to and Eq.~(\ref{Eq_C_cur_tor}), the signed curvature of plane curve is given by
\begin{equation}
\kappa(\lambda) = \frac{x^{\prime} y^{\prime \prime}-y^{\prime} x^{\prime \prime}}{\left(x^{\prime 2}+y^{\prime 2}\right)^{3 / 2}}.
\end{equation}
Consider $\mathbf{r}(\lambda)=(\sin 2\lambda, 3.5\sin\lambda,0 )$ with $\lambda \in [0,2\pi]$ as an example, we have $\kappa(\lambda) = (10.5 \sin\lambda + 3.5 \sin 3\lambda)/||\mathbf{r}^{\prime}||^{3}$ with $||\mathbf{r}^{\prime}|| = (12.25\cos^2\lambda + 4\cos^2 2\lambda)^{1/2}$. To obtain the corresponding pulse, we perform a numerical variable transformation to obtain the curvature in terms of curve length variable $t$, i.e., $\Omega(t) = \kappa(t)$. Systematic construction of the first and second-order robust control pulses from analytical plane curves is discussed in the next section.

To demonstrate a space curve with curvature zeros, we take a figure 8 space curve shown in Fig.~\ref{Fig_spacecurve}(b) $\mathbf{r}(\lambda)=(\sin 2\lambda, 3.5\sin\lambda, 3.5\sin\lambda )$ with $\lambda \in [0,2\pi]$ for example. We have $\sin\psi = \sin\theta = 3.5\cos \lambda / ||\mathbf{r}^{\prime}||$ and $\tan\psi = - 2\cos 2\lambda / ||\mathbf{r}^{\prime}||$ with $||\mathbf{r}^{\prime}|| = (24.5 \cos^2\lambda + 4 \cos^2 2 \lambda)^{1/2}$. At the point of curvature singularity $t=\pi$, $\tan\phi$ is still well-defined by taking the limit of Eq.~(\ref{Eq_C_phi}). The continuous tangent and normal vector are illustrated using arrows on the space curve. The conventional curvature and signed curvature are defined by $\| \mathbf{r}^{\prime} \times \mathbf{r}^{\prime\prime} \| / \|\mathbf{r}^{\prime}\|^{3}$ and Eq.~(\ref{Eq_C_cur_tor}) in terms of curve coordinates and the continuous Frenet vectors respectively and are plotted in Fig.~\ref{Fig_spacecurve}(b) for comparison. The curvature zero at $t=\pi$ will lead to a singularity and discontinuity of the normal vector and torsion in the framework of standard differential geometry of space curves, while their counterparts obtained by the Frenet angles is continuous and singularity free. The corresponding robust pulses are given by the signed curvature and singular free torsion ($\tau(\lambda)=0$ for this example) obtained from the continuous Frenet vectors (Eq.~(\ref{Eq_C_cur_tor})). The robust pulse performs an identity operation, and its robustness against detuning noise is presented in Fig.~\ref{Fig_spacecurve}(b).

\section{Robust Pulses from Analytical Construction of Plane Curves}
\label{AppendixD}

\begin{figure*}[th]
\centering
\includegraphics[width=1\textwidth]{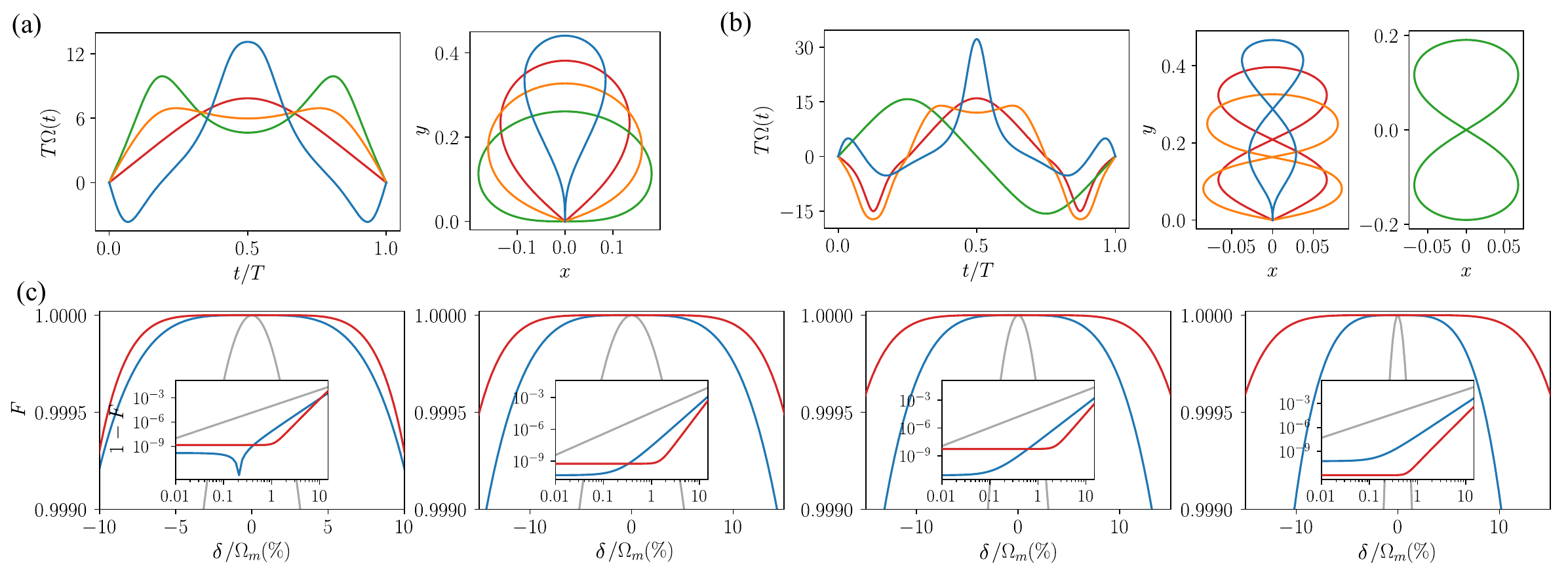}
\caption{(a) First order robust control pulses $\{ R_{1^{\prime};\bot}^{\pi},R_{1^{\prime};\bot}^{7\pi/4},R_{1^{\prime};\bot}^{5\pi/2},R_{1^{\prime};\bot}^{2\pi} \}$ (left) obtained from analytical plane curves (right). 
(b) Second order robust control pulses $\{ R_{2;\bot}^{\pi},R_{2;\bot}^{7\pi/4},R_{2;\bot}^{5\pi/2},R_{2;\bot}^{2\pi} \}$ (left) obtained from analytical plane curves (right). 
(c) Fidelity of four single-qubit gates $\{ X_{\pi},X_{\pi/4},X_{\pi/2},X_{2\pi} \}$ (left to right) realized by the second order, first order RCPs shown in (b)(a) and their cosine pulse counterpart (red, blue, grey) against detuning noise. Insets: Infidelity as a function of the relative noise strength for each gate in log-log scale.
}
\label{Fig_ana_pulse}
\end{figure*}

In this section, we construct a set of first and second-order RCPs against longitudinal error from analytical plane curves that are closed and with zero net areas, which are z-error curves of the corresponding dynamics satisfying the first and second-order robustness conditions.

First of all, a series of first-order RCPs can be generated from the modified half lemniscates of Bernoulli used in \cite{zeng2018general} 
\begin{equation}
\begin{array}{c}
x_{1}(\lambda )=\frac{\alpha \sin (2\lambda )}{3+\cos (2\lambda )} \\ 
y_{1}(\lambda )=\frac{2\sin (\lambda )}{3+\cos (2\lambda )}.
\end{array}%
\label{Eq_D_curve1}
\end{equation}%
When $0\leq \lambda \leq \pi $, it becomes closed curve that subtends an
angle $\theta =\pi -2\arctan (\frac{1}{\alpha })$ at the origin. 
The rotation angle of the corresponding pulse is given by the total curvature or total winding angle of the tangent vector of the curve $\phi=\int_{0}^{\Lambda}dt \kappa(\lambda) = 2\pi-2\arctan(\frac{1}{\alpha})$.
Thus, through tuning $\alpha$, one can obtain the curves for first order
RCPs with rotational angle $\pi<\phi<2\pi$. In addition, the curve for $%
\pi $ pulse can be obtained by multiplying $x_{1}(\lambda)$ by a power of $%
\sin(\lambda)$ to diminish the angle subtended at the origin and guarantee the
curvature starts and ends at zero. A first-order $2\pi$ pulse can be
constructed from a modified circle.

We construct four plane curves for first order RCPs $\{ R_{1^{\prime};\bot}^{\pi},R_{1^{\prime};\bot}^{7\pi/4},R_{1^{\prime};\bot}^{5\pi/2},R_{1^{\prime};\bot}^{2\pi} \}$. The analytical expressions of the plane curves are listed in Table.~\ref{table2}, and the corresponding RCPs are obtained by calculating their curvature in the unit-speed parametrization. The four first-order RCPs and the corresponding plane curves are shown in Fig.~\ref{Fig_ana_pulse}(a).

We then present a piecewise construction of the plane curves for second order
RCPs with rotation angle $0<\phi <\pi $. The basic constituents of the
composite curve are the aforementioned modified Bernoulli half lemniscates $\{x_{1}(\lambda
),y_{1}(\lambda )\}$ and the sinusoidal curve
\begin{equation}
\begin{array}{c}
x_{2}(\lambda )=\alpha \sin (2\lambda ) \\ 
y_{2}(\lambda )=2\lambda .
\end{array}
\label{Eq_D_curve2}
\end{equation}%
The equation of the composite curve is 
\begin{equation}
\left\{ 
\begin{array}{l}
x(\lambda )=-\beta x_{2}(\lambda ),\text{}y(\lambda )=\beta y_{2}(\lambda
),\quad 0\leq \lambda <\frac{\pi }{2} \\ 
x(\lambda )=x_{1}(\lambda -\frac{\pi }{2}),\text{ }y(\lambda )=\beta y_{2}(%
\frac{\pi }{2})+y_{1}(\lambda -\frac{\pi }{2}), \\ 
\multicolumn{1}{r}{\frac{\pi }{2}\leq \lambda <\frac{3\pi }{2}} \\ 
x(\lambda )=\beta x_{2}(2\pi -\lambda ),\text{ }y(\lambda )=\beta
y_{2}(2\pi -\lambda ), \\ 
\multicolumn{1}{r}{\frac{3\pi }{2}\leq \lambda \leq 2\pi }%
\end{array}%
\right. 
\label{Eq_D_comp_curve}
\end{equation}%
where $\alpha $ and $\beta $ are two parameters determined by the
target rotation angle $\phi =2\arctan (\frac{1}{\alpha })$ and the zero
net-area condition $\int_{0}^{2\pi }(y^{\prime }x-x^{\prime }y)d\lambda =0$.

We construct RCPs $R_{2;\bot}^{\pi/4}$ and $R_{2;\bot}^{\pi/2}$ from the composite curve of Eq.~(\ref{Eq_D_comp_curve}), the RCP $R_{2;\bot}^{\pi}$ is obtained by modifying the curves for $R_{1^{\prime};\bot}^{\pi}$ and $R_{2;\bot}^{2\pi}$ is generated from the curves Eq.~(\ref{Eq_D_curve1}) with $0\le \lambda \le 2\pi$. The four plane curves and the corresponding second-order RCPs are shown in Fig.~\ref{Fig_ana_pulse}(b), and the analytical expressions for the four constructed curves are listed in Table~\ref{table2}.

Fig.~\ref{Fig_ana_pulse}(c) shows the robustness of the single qubit gate $\{ X_{\pi},X_{\pi/4},X_{\pi/2},X_{2\pi} \}$ 
performed by the first and second-order RCPs mentioned above against detuning noise. All of the RCPs exhibit robust infidelity plateau with values much smaller than that of the cosine pulse within the noise region from $0.01\%$ to $1\%$.

\begin{table*} %
\setlength{\tabcolsep}{20pt}
\renewcommand{\arraystretch}{1.4}
\begin{tabular}{ll}
\hline\hline
RCPs &  Curve Functions \\ \hline
$R_{1^{\prime};\bot}^{3\pi/2}$ & $x(\lambda) = x_{1}(\lambda)$ \quad$y(\lambda) =
y_{1}(\lambda)$, \quad$\alpha= 1$, \quad $0 \le\lambda\le\pi$ \\ \hline

\multirow{4}{3em}
{$R_{2;\bot}^{\pi/2}$} & $x(\lambda) = -\beta x_{2}(\lambda)$ \quad$y(\lambda) = \beta
y_{2}(\lambda)$, \quad$0 \le\lambda< \frac{\pi}{2}$  \\ 
        & $x(\lambda) = x_{1}(\lambda-\frac{\pi}{2})$ \quad$y(\lambda) = \beta
y_{2}(\frac{\pi}{2}) + y_{1}(\lambda-\frac{\pi}{2})$, \quad$\frac{\pi}{2}
\le\lambda< \frac{3\pi}{2}$ \\
& $x(\lambda) = \beta x_{2}(2\pi- \lambda)$ \quad$y(\lambda) = \beta
y_{2}(2\pi- \lambda)$, \quad$\frac{3\pi}{2} \le\lambda\le2\pi$\\
& $\alpha= 1$ \quad$\beta= 0.3535534$   \\ \hline

$R_{1^{\prime};\bot}^{7\pi/4}$ &   $x_{1^{\prime}}(\lambda) = x_{1}(\lambda)$ \quad $y_{1^{\prime}}(\lambda) =
y_{1}(\lambda)(-0.3\lambda (\lambda - \pi) + 1)$, \quad $\alpha= 1$, \quad $0 \le\lambda\le\pi$ \\ \hline

\multirow{4}{3em}
{$R_{2;\bot}^{\pi/4}$} & $x(\lambda) = -\beta x_{2}(\lambda)$ \quad $y(\lambda) = \beta
y_{2}(\lambda)$, \quad $0 \le\lambda< \frac{\pi}{2}$  \\
& $x(\lambda) = x_{1^{\prime}}(\lambda-\frac{\pi}{2})$ \quad $ y(\lambda) = \beta
y_{2}(\frac{\pi}{2}) + y_{1^{\prime}}(\lambda-\frac{\pi}{2})$, \quad $\frac{\pi}{2}
\le\lambda< \frac{3\pi}{2}$ \\
& $x(\lambda) = \beta x_{2}(2\pi- \lambda)$ \quad$y(\lambda) = \beta
y_{2}(2\pi- \lambda)$, \quad$\frac{3\pi}{2} \le\lambda\le2\pi$\\
& $\alpha= 2.4142136$ \quad $\beta= 0.4801245$   \\ \hline

$R_{1^{\prime};\bot}^{\pi}$ &  $x_{1^{\prime}}(\lambda) = x_{1}(\lambda)\sin^{2}(\lambda)$ \quad $y_{1^{\prime}}(\lambda) = y_{1}(\lambda)$, \quad $\alpha= 0.72$, \quad $0 \le\lambda\le\pi$  \\  \hline

\multirow{2}{3em}
{$R_{2;\bot}^{\pi}$} & $x(\lambda) = x_{1^{\prime}}(\lambda)(\lambda- (\frac{\pi}{2}-b))(\lambda- (\frac{\pi}{2}+b))$ \quad $y(\lambda) = 0.25y_{1^{\prime}}(\lambda)$  \\
& $\alpha= -0.3$, $b = 0.6100818$ \quad $0 \le\lambda\le\pi$ \\  \hline

$R_{1^{\prime};\bot}^{2\pi}$ & $x(\lambda) = \frac{ 2.4 \sin(2\lambda+Pi) }{2 + \cos 2\lambda } $ \quad $ y(\lambda) = \frac{ \cos(2\lambda + \pi) + 1 }{2 + \cos 2\lambda )} \sin(\lambda+\pi) $ \\  \hline

$R_{2;\bot}^{2\pi}$ & $x(\lambda) = x_{1}(\lambda)$ \quad$y(\lambda) =
y_{1}(\lambda)$, \quad$\alpha= 1$, \quad $0 \le\lambda\le 2\pi$  \\   \hline\hline

\end{tabular}
\caption{Plane curve functions for RCPs. Curve ansatz $\{ x_1, y_1\}$ and $\{ x_2, y_2\}$ given by Eq.~(\ref{Eq_D_curve1}) and Eq.~(\ref{Eq_D_curve2}) are used. For the curve functions of RCPs, $R_{1^{\prime};\bot}^{7\pi/4}$ and $R_{1^{\prime};\bot}^{\pi}$, additional modifications are made to smoothen the resulting pulses, and the modified curve functions (denoted by $\{ x_{1^{\prime}}, y_{1^{\prime}}\}$) are used in constructing the plane curves for $R_{2;\bot}^{7\pi/4}$ and $R_{2;\bot}^{\pi}$ pulses. }
\label{table2}
\end{table*}

\section{Supplementary for numerical results}
\label{AppendixB}

\begin{figure}[t]
	\centering
	\includegraphics[scale=0.5]{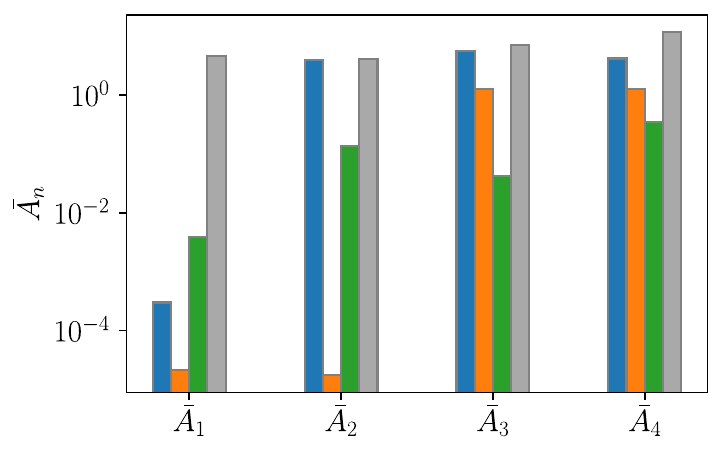}
	\caption{ Numerical certification of robust pulses. The first to fourth order rescaled Magnus error coefficients for the $X_{\pi/4}$ gate evolution produced by $R_{1;\bot }^{\pi/4}$, $R_{2;\bot}^{\pi/4}$, $R_{\text{ex};\bot}^{\pi/4}$ and cosine pulse (blue, orange, green, grey).    }
	\label{Fig_robustness_order}
\end{figure}

As an additional assessment of the robustness of our RCPs presented in Sec.~(\ref{Sec_protocol}), we calculate the Magnus expansion coefficients up to the fourth order numerically. The rescaled Magnus coefficients $\bar{A}_n = 10^{-n}||A_n||$ of the $X_{\pi/4}$ gate evolution produced by the $R_{1;\bot }^{\pi/4}$, $R_{2;\bot}^{\pi/4}$, $R_{\text{ex};\bot}^{\pi/4}$ and cosine pulse are plotted in Fig.~\ref{Fig_robustness_order} for comparison, where $R_{1;\bot }^{\pi/4}$ and $R_{\text{ex};\bot}^{\pi/4}$ are RCPs obtained by our pulse generation protocol and $R_{2;\bot}^{\pi/4}$ is an RCP with second order robustness constructed from the analytical plane curve in Appendix~\ref{AppendixD}.

Compared with cosine pulse, $R_{1;\bot }^{\pi/4}$ and $R_{2;\bot}^{\pi/4}$ have small Magnus coefficients up to first and second order respectively, while all four coefficients for $R_{\text{ex};\bot}^{\pi/4}$ pulse are significantly suppressed, indicating its higher robustness.

\begin{figure}[t]
	\centering
	\includegraphics[width=1\columnwidth]{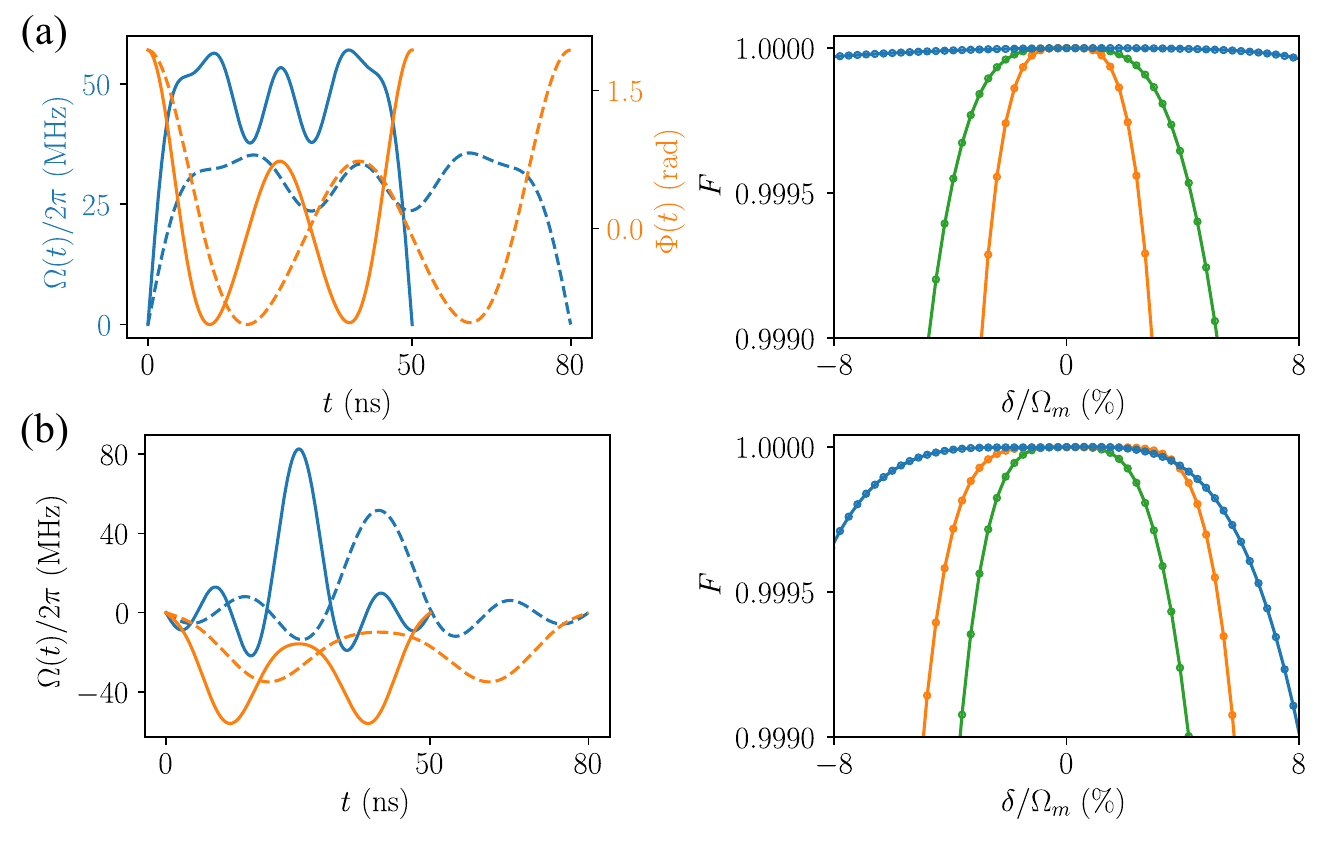}
	\caption{(a) Left: the $R_{1;\text{all}}^{3\pi/2}$ pulses with 50 ns duration (solid) and the rescaled pulses with 80 ns duration (dashed). Right: fidelity of the corresponding $Y_{3\pi/2}$ gate generated by the 50 ns pulses (solid) as a function of the relative noise strengths of $\delta_z$, $\delta_x$, and $\delta_y$ (blue, orange, and green) respectively, and the gate fidelity by 80 ns pulses as a function of the same noise strengths (circle).
(b) Left: the $R_{1;\text{all}}^{\pi}$ pulses in the X-Y driven scheme with 50 ns duration (solid) and the rescaled pulses with 80 ns duration (dashed). Right: fidelity of the corresponding $Y_{3\pi/2}$ gate generated by the 50 ns pulses (solid) as a function of the relative noise strengths of $\delta_z$, $\delta_x$, and $\delta_y$ (blue, orange, and green) respectively, and the gate fidelity by 80 ns pulses as a function of the same noise strengths (circle).    }
	\label{Fig_pulse_rescale}
\end{figure}

Our robust control pulses are suitable for different gate times, and the pulse rescaling in the time domain does not change their robustness since the substitution $t \rightarrow \alpha t$, $\Omega \rightarrow \Omega/\alpha $ does not change Eq.~(\ref{Eq_z_correspondence2}) and thus maintains the geometric correspondence between pulses and their error curves. 
Here we present a $50$ ns pulse for $3\pi/2$ rotation around the Y axis by adding a $\pi/2$ phase constant to the RCP $R_{1;\text{all}}^{3\pi/2}$ presented in the main text and rescale it to an $80$ ns pulse. It maintains robustness in three axes with error Hamiltonian in the form of Eq.~(\ref{Eq_GendeltaH}). Fig.~\ref{Fig_pulse_rescale}(a) shows the rescaled pulse and the gate fidelity of the $Y_{3\pi/2}$ gate generated by the two pulses. The coinciding fidelity values as a function of the relative noise strength demonstrate their same robustness.

In addition, our pulse generation protocol is compatible with a direct x-y control scheme, in which the noiseless Hamiltonian is of the form $H_0(t) = \Omega_x(t)/2\sigma_x + \Omega_y(t)/2\sigma_y$. Here we present a $50$ ns pulse for $\pi$ rotation around the X axis (denoted by $R_{1;\text{all}}^{\pi}$) that has robustness in three axes and rescale it to a $100$ ns pulse. Fig.~\ref{Fig_pulse_rescale}(b) shows the rescaled pulse and the gate fidelity of the $X_{\pi}$ gate generated by the two pulses.

The parameters of all RCPs are presented in the main text, and this section is listed in Table~\ref{table1}.

\begin{table*} %
\setlength{\tabcolsep}{20pt}
\renewcommand{\arraystretch}{1.3}
\begin{tabular}{llll}
\hline\hline
RCPs &  & $a$ & $\phi$ \\ \hline
$R_{1;\bot }^{\pi}$ & $\Omega$ & $[0.010,-0.259,-0.033]$ & $[-0.015,-0.038]$ \\ 
$R_{1;\bot }^{7\pi/4}$ & $\Omega$ & $[0.223,0.134,0.076]$ & $[0.001,-0.020]$ \\ 
$R_{1;\bot }^{5\pi/2}$ & $\Omega$ & [0.349,0.307] & [-0.003] \\ 
$R_{1;\bot }^{2\pi}$ & $\Omega$ & [0.258,0.183] & [0] \\  

$R_{\text{ex};\bot}^{\pi}$ & $\Omega$ & $[-0.328,-1.014,-1.195,-0.304]$ & $[-0.003,-0.003,-0.008]$ \\ 
$R_{\text{ex};\bot}^{9\pi/4}$ & $\Omega$ & $[0.147,-0.089,-0.613,-0.161]$ & $[-0.123,-0.061,-0.073]$ \\ 
$R_{\text{ex};\bot}^{5\pi/2}$ & $\Omega$ & $[0.241,0.084,-0.482,-0.036]$ & $[-0.036,0.014,0.107]$ \\ 
$R_{\text{ex};\bot}^{2\pi}$ & $\Omega$ & $[0.042,-0.290,-0.765,-0.274]$ & $[0.003,0.003,0.003]$ \\ 
\multirow{2}{3em}{$R_{1;\text{all}}^{3\pi/2}$} & $\Omega$ & $[0.624,0.484,0.193,0.070,0.073]$ & $[0.005,0.013,0.003,-0.070]$ \\
        & $\Phi$ & $[0.083,0.362,1.174,0.237,0.074]$ & $[0.022,0.017,0.011,-0.037]$ \\
\multirow{2}{3em}{$R_{1;\text{all}}^{\pi}$} & $\Omega_x$ & $[0.007,-0.236,0.032,-0.250]$ & $[0.008,-0.601,-0.029]$ \\
        & $\Omega_y$ & $[-0.327,-0.127,0.167,0.066]$ & $[0.035,-0.079,-0.096]$ \\   \hline\hline
\end{tabular}
\caption{Parameters for RCPs with gate time $T=50$ ns and amplitude unit GHz.}
\label{table1}
\end{table*}
\end{appendix}

%

\end{document}